\documentclass[
    aps,
    prb,
    preprint,
    superscriptaddress
]{revtex4-2}

\usepackage{float}
\usepackage{graphicx}
\usepackage{bm}
\usepackage{amsmath, amssymb} % Works perfectly with RevTeX
\usepackage{booktabs}
\usepackage{xcolor}
\usepackage{hyperref}
\usepackage{tikz} 
\usepackage[abs]{overpic} 
\usepackage{etoolbox} 
\usepackage{setspace}
\definecolor{pink}{rgb}{0.78,0.08,0.52}

\hypersetup{
    colorlinks=true,
    linkcolor=magenta,
    citecolor=blue,
    filecolor=magenta,
    urlcolor=violet,
}

\begin{document}

\title{Stability of Continuous Time Quantum Walks in Complex Networks}
\author{Adithya L J}
\email{adiethu@gmail.com}
\affiliation{Department of Physics, Indian Institute of Science Education and Research Pune, 411008, India}

\author{Johannes Nokkala}
\email{jsinok@utu.fi}
\affiliation{Department of Physics and Astronomy, University of Turku, FI-20014 Turun yliopisto, Finland}
\affiliation{Turku Collegium for Science, Medicine and Technology, University of Turku, Turku, Finland}

\author{Jyrki Piilo}
\email{jyrki.piilo@utu.fi}
\affiliation{Department of Physics and Astronomy, University of Turku, FI-20014 Turun yliopisto, Finland}

\author{Chandrakala Meena}
\email{meenachandrakala@gmail.com} 
\affiliation{Department of Physics, Indian Institute of Science Education and Research Pune, 411008, India}

\date{\today}
\begin{abstract}
    We investigate the stability of continuous-time quantum walks (CTQW) across cycle, complete, star, Erdős–Rényi, small-world, and scale-free topologies under energy-based intrinsic decoherence, node-based Haken–Strobl noise, and edge-based quantum stochastic walk (QSW) decoherence. Defining stability as the preservation of quantum properties, we characterize it using node probabilities, $\ell_1$-norm of coherence, fidelity, quantum-classical distance, and von Neumann entropy. Our results show that intrinsic decoherence preserves coherence longest, while QSW causes the most rapid decay. Stability rankings varies and depend on the decoherence types, network structure, and properties of node where walker is initialized specifically in heterogeneous networks. Dense connected network like complete and heterogeneous networks, for instance star, and scale-free are stable under Haken–Strobl noise but become uniquely fragile under QSW when initialized on high-degree nodes. However, these same networks, due to their inherent localization, exhibit lower coherence in the noiseless regime, highlighting a fundamental trade-off between localization and coherence. Furthermore, the centrality of the initialization node has a pronounced impact on relaxation time and stability measures, underscoring the critical role of local topological features in quantum dynamics.

    \vspace{10pt}

\end{abstract}

\maketitle

\section{Introduction}
Quantum walks (QWs) are quantum analogs of classical random walks, have emerged as a powerful framework for exploring a wide range of quantum phenomena, including quantum computation, coherent transport, and the dynamical behavior of complex quantum systems~\cite{aharonov1993quantum,ambainis2003quantum,konno2008quantum}. Unlike their classical counterparts, QWs exploit quantum superposition and interference, enabling the walker to explore graphs more efficiently and, in some cases, achieve quantum speedups in algorithmic tasks~\cite{childs2004spatial}.

QWs can be broadly categorized into two types: discrete-time quantum walks (DTQW)~\cite{aharonov1993quantum} and continuous-time quantum walks (CTQW)~\cite{farhi1998quantum}. DTQWs evolve via alternating applications of a coin operator and a conditional shift, allowing fine-grained control over the walker’s trajectory~\cite{jayakody2023revisiting}. In contrast, CTQW evolve under a time-independent Hamiltonian defined by the graph structure, without requiring an internal coin degree of freedom. This formulation is particularly well-suited for modeling coherent transport in structured networks and has applications in quantum search~\cite{childs2004spatial,chakraborty2016spatial,apers2022quadratic}, quantum state transfer~\cite{coutinho2025peak}, entanglement routing~\cite{sauglam2023entanglement}, and energy transport in photosynthetic complexes~\cite{mohseni2008environment,plenio2008dephasing,alexander2025quantum}. However, real-world implementations of QWs are inevitably affected by decoherence arising from environmental interactions and  system imperfections. These effects disrupt unitary evolution, suppress quantum coherence, and reduce transport efficiency and computational fidelity~\cite{ren2009decoherence,allegra2016global,bressanini2022decoherence}. Interestingly, in some cases decoherence can also assist transport~\cite{caruso2009highly,plenio2008dephasing,allegra2016global}. Understanding how decoherence influences QWs is therefore critical for the development of robust quantum technologies.

Decoherence in open quantum systems is typically continuous rather than discrete~\cite{breuer2002theory}, our study focuses on CTQW subjected to various continuous decoherence mechanisms. Common modeling approaches include intrinsic decoherence, modeled as dephasing in the energy eigenbasis via the Milburn equation~\cite{milburn1991intrinsic}; position-basis decoherence, described by the Haken–Strobl master equation~\cite{haken1973exactly}; and quantum stochastic walks (QSW), which interpolate between quantum and classical regimes~\cite{QSW}. 
These frameworks thus enable us to explore different facets of open quantum dynamics and their consequences for quantum transport.

The evolution of CTQW is influenced by both the form of decoherence and the structure of the underlying graph~\cite{bressanini2022decoherence,strauch2009reexamination}. Earlier studies have explored CTQW dynamics on one-dimensional lattices to examine quantum transport~\cite{siloi2017noisy} and quantum correlations~\cite{benedetti2012quantum,PhysRevLett.127.100406}, while more recent efforts have extended this work to symmetric~\cite{bressanini2022decoherence} and complex networks~\cite{rossi2014node,mulken2011continuous,tsomokos2011quantum,malmi2022spatial}.

In this work, we broaden the scope by considering a diverse set of network topologies, and evaluating their behavior under multiple decoherence models. Our primary objective is to understand how the interplay between network structure and decoherence governs the behavior and stability of CTQW.

Motivated by analogous findings in classical network theory, where the stability of dynamical states is often determined by the interplay between topology and dynamics~\cite{boccaletti2006complex,mulken2011continuous,meena2023emergent}, so in this study we seek to uncover similar principles within the context of quantum systems. 
Specifically, we investigate how the CTQW dynamics evolves under various decoherence mechanisms across different network architectures. Here, we define quantum stability as the ability of a system to retain its quantum features, such as coherence, over time, despite the presence of decoherence.

It is well established that node's centralities, such as degree and closeness, play a critical role in determining the resilience of classical networks~\cite{rungta2018identifying,meena2017threshold}. Recent studies indicate that these structural attributes also significantly influence quantum dynamics~\cite{böttcher2021classical}.
Identifying which topological features help mitigate or exacerbate decoherence is essential for designing stable and efficient quantum systems. Such insights are crucial for advancing robust quantum communication protocols, quantum devices, and fault-tolerant quantum architectures~\cite{bose2007quantum}. Motivated by this, our study also investigates how node centralities affect quantum stability, with a particular focus on heterogeneous networks, where centrality disparities are pronounced. To quantify the stability of CTQW under different decoherence models, we employ several metrics used in the analysis of open quantum systems, including quantum-classical distance, von Neumann entropy, $\ell_1$-norm of coherence, and fidelity. These measures provide a multifaceted view of how far the system deviates from ideal quantum behavior over time. For example, the quantum-classical distance has been used to study how CTQW diverge from quantum behavior under decoherence in simple network topologies such as complete, cycle, and star graphs~\cite{bressanini2022decoherence}.

Building on this foundation, we extend the analysis to more complex graphs and a broader range of decoherence mechanisms. Specifically, this study investigates how different forms of decoherence affect the resilience of CTQW across both simple and complex network structures. Our results provide insights into open quantum dynamics on networks and offer guidance for designing quantum systems that maintain coherence in the presence of environmental disturbances.

We begin Sec.~\ref{section1} by detailing the dynamics of CTQW in the presence of various decoherence mechanisms. In Sec.~\ref{section2}, we provide an overview of the network structures considered in our study. Sec.~\ref{section3} introduces the quantitative measures employed to assess the stability of CTQW across different network structures. Results are presented in Sec.~\ref{section4}, where we analyze how decoherence impacts quantum properties on both simple and complex networks. Finally, Sec.~\ref{section5} concludes the paper with a summary of our key results and their broader implications.

\section{Theoretical Framework}
\subsection{Dynamical framework of CTQW on networks} \label{section1}
In a CTQW on a network, the walker is initially localized at a specific node and gradually explores the network as time evolves. The evolution of the quantum state is governed by a Hamiltonian, typically chosen to be the graph Laplacian of the underlying network.

The Laplacian matrix is defined as $L=D-A$, where $D$ is the diagonal degree matrix with the degree (i.e., number of connections) of each node as its diagonal elements, and $A$ is the adjacency matrix, with elements $A_{ij}=1$ if node $i$ and $j$ are connected and $A_{ij}=0$ otherwise. 

We investigate the impact of three distinct types of decoherence on CTQW: (i) intrinsic decoherence, modeled by the Milburn equation~\cite{milburn1991intrinsic}, (ii) position-basis decoherence, described by the Haken–Strobl master equation~\cite{haken1973exactly}, and (iii) quantum stochastic walks, as formulated in~\cite{caruso2014universally}. Each of these models introduces a unique form of non-unitary evolution, enabling us to explore different facets of quantum system dynamics under decoherence on networks.

In the absence of decoherence, that is, ideal noiseless conditions, the quantum walker evolves coherently, governed by unitary dynamics. This evolution is described by the Liouville von Neumann equation~\cite{breuer2002theory}:

\begin{equation}
    \frac{d\rho(t)}{dt} = -i[H, \rho(t)].
    \label{noiseless}
\end{equation}
Where $\rho(t)$ denotes the density matrix of the system at time $t$, and $H$ is the system's Hamiltonian, typically chosen as the graph Laplacian $L$ in the context of network-based quantum walks.

To model intrinsic decoherence, we take Milburn formalism~\cite{milburn1991intrinsic}, which introduces non-unitary dynamics by modifying the noiseless evolution Eq.~\eqref{noiseless} as follows:
\begin{equation}
\frac{d\rho(t)}{dt} = -i [H, \rho(t)] - \frac{\gamma}{2} [ H, [H, \rho(t)] ].
\label{intrinsic_decoherence}
\end{equation}
Here, 
$\gamma$ represents the decoherence rate. This model captures intrinsic decoherence, which reflects the system’s internal loss of coherence, independent of external environmental interactions. The double-commuter term in Eq.~\eqref{intrinsic_decoherence} models decoherence in the energy basis.

The next model we consider is based on the Haken-Strobl master equation, which is commonly employed to describe dephasing noise on the position basis. It takes the following form~\cite{haken1973exactly}:

\begin{equation}
        \frac{d\rho(t)}{dt} = -i[H, \rho(t)] + \gamma\sum_k \left( P_k \rho(t) P_k^\dagger - \frac{1}{2} \{ P_k^\dagger P_k, \rho(t) \} \right);
        \label{eq.HS}
\end{equation}
  $P_k=|k\rangle \langle k|$ are projectors on the position basis. 
This equation models decoherence by coupling the system to a bath that causes loss of coherence in the position(or node) basis. As the projectors $P_k$ act on individual nodes, this is a node based decoherence model.
  
The last decoherence model is Quantum Stochastic Walk (QSW)~\cite{caruso2014universally} that interpolates between pure quantum and classical behavior. It is described by the equation as follows~\cite{QSW}:
\begin{equation}
\begin{split}
\frac{d\rho(t)}{dt} =\, & -(1 - p)i[H, \rho(t)] \\
& + p \sum_{kj} \left( P_{kj} \rho(t) P_{kj}^\dagger 
- \frac{1}{2} \left\{ P_{kj}^\dagger P_{kj}, \rho(t) \right\} \right);
\end{split}
\label{eq.QSW}
\end{equation}$P_{kj}=L_{kj}|k\rangle \langle j|$, this jump operators are dependent on connections(edges) between nodes making it a edge based decoherence model. The parameter 
$p$ controls the interpolation between unitary evolution and stochastic noise. 

\subsection{Network Topologies}\label{section2}

We consider a quantum walker evolving under CTQW dynamics, where the walker is initially localized at a single node in the network. This approach is applied consistently across all the networks that we consider in this study. The rationale behind this choice is to minimize the fidelity between the quantum-evolved state and the classically evolved state, as discussed in prior studies on CTQW~\cite{gualtieri2020quantum}. As the system evolves, the walker’s state becomes a quantum superposition over multiple nodes, with the probability of finding the walker at a given node determined by the diagonal terms of the density matrix.

As a baseline for comparison with more complex networks, we begin our analysis with three simple graph structures: complete, cycle, and star graphs, each have a size of node number $N=10$. In a complete graph, every node is connected to all other nodes, resulting in a homogeneous network, where each node has degree of $N-1$. The cycle graph is also a homogeneous network, where each node is connected only to its two nearest neighbors, yielding a uniform degree of $2$. In contrast, the star graph exhibits a highly heterogeneous structure in which the central hub node has degree $N-1$, while all peripheral (edge) nodes are connected only to the center node and hence have degree $1$.
  
We consider three types of complex network topologies for our simulations: Barabási–Albert scale-free networks~\cite{barabasi1999emergence}, Erdős–Rényi random networks~\cite{erdos1960evolution}, and Watts–Strogatz small-world networks~\cite{watts1998collective}. These network models differ significantly in their structural characteristics due to distinct connectivity patterns. For example, Erdős–Rényi networks~\cite{erdos1960evolution} are homogeneous, with a binomial degree distribution tightly centered around the average degree, making the nodes statistically similar in connectivity. In contrast, Barabási–Albert networks~\cite{barabasi1999emergence} are scale-free, characterized by a power-law degree distribution. This results in a few highly connected hub nodes, introducing significant degree heterogeneity into the network. The Watts–Strogatz model~\cite{watts1998collective} generates small-world networks with high clustering and short average path lengths. Starting from a regular ring lattice where each node is connected to its four nearest neighbors, a small fraction of the edges is randomly rewired with probability $p=0.1$, introducing long-range shortcuts while maintaining local structure~\cite{mulken2007quantum}. This process disrupts the periodicity of the lattice, creating a network that is partially random but still relatively homogeneous in degree distribution. 
Scale-free networks exhibit strong degree heterogeneity due to the emergence of hubs, while Erdős–Rényi and small-world networks remain relatively homogeneous in their degree distributions. For all simulations involving these network topologies, we consider networks of size 
$N=100$ nodes, each with an average degree of $4$.

\subsection{Stability Measures for CTQW on networks}\label{section3}
\begin{figure*}[htbp]
    \centering
    \includegraphics[width=1\textwidth]{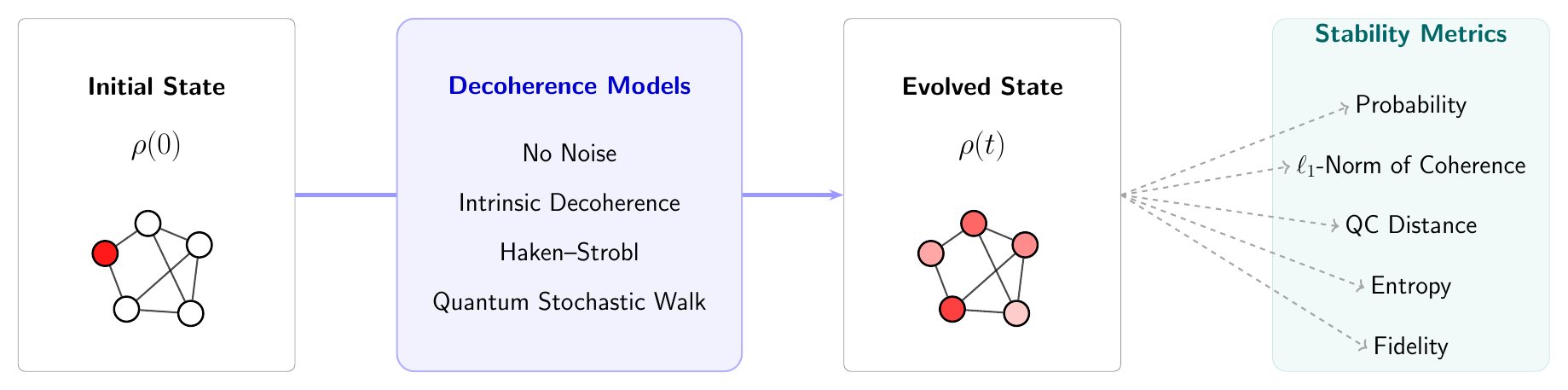}
  \caption{\textbf{Schematic diagram for determining quantum stability}. To determine the stability of the quantum walker which is initially localized on a specific node of a network (red) and as it evolves over time, its probability distribution spreads across the network, visualized through nodes of varying red opacity. The evolution of the walker depends on the type of decoherence and the network topology. The stability analysis is performed using probability distribution, $\ell_1$-norm of coherence, fidelity with initial state, von Neumann entropy, and quantum-classical distance.}
    \label{fig:illustration}
\end{figure*}
To investigate the stability of CTQW on networks under decoherence, we analyze the occupation probability of the walker across nodes, the $\ell_1$-norm of coherence, the fidelity with respect to the initial state, the von Neumann entropy, and the quantum–classical distance, as illustrated in Fig.~\ref{fig:illustration}. These metrics are complementary, each capturing distinct aspects of the system's behavior and collectively providing a comprehensive picture of stability under decoherence. The following subsections provide brief descriptions of each of these measures.

\subsubsection{Occupation Probability}
We analyze the node occupation probability and its long-time behavior to understand the influence of decoherence on CTQW. The occupation probabilities correspond to the diagonal elements of the density matrix. In the presence of decoherence, the occupation probability evolves toward a steady state over time. 
The rate at which the system approaches its steady state serves as an informative measure of the CTQW's stability in the presence of decoherence.
Slower convergence reflects greater stability of the CTQW under specific decoherence. Moreover, when the occupation probabilities approach uniformity across nodes, the system tends toward classical-like behavior, signaling a transition from quantum to classical~\cite{mulken2011continuous}.
\subsubsection{$\ell_1$-norm of coherence}
$\ell_1$-norm of coherence is the sum of the absolute values
of all off-diagonal elements of the density matrix~\cite{baumgratz2014quantifying}. 
\begin{equation}
C_{\ell_1}(\rho(t)) = \sum_{i \neq j} |\rho_{ij}(t)|.
\end{equation}Decay of the $\ell_1$-norm of coherence toward zero signifies the gradual loss of coherence, indicating that the system is undergoing classicalization under the given decoherence mechanism. In contrast, if a non-negligible level of coherence is maintained over time, the CTQW retains its quantum character and can be considered as stable.

\subsubsection{Fidelity}

Fidelity quantifies the degree of overlap between the initial quantum state and the state evolved under decoherence~\cite{jozsa1994fidelity,raginsky2001fidelity,gilchrist2005distance,chen2018simulating}. It measures how closely the evolved state overlaps to the initial state. Given the density matrix $\rho(t)$ at time $t$ and the initial density matrix $\rho(t)$, the fidelity $F$ is defined:

\begin{equation}
    F(\rho_0, \rho(t)) = \left[ \text{Tr} \left( \sqrt{\sqrt{\rho_0} \rho(t) \sqrt{\rho_0}} \right) \right]^2.
\end{equation}Higher fidelity indicates that the quantum state retains more of its initial state over time, reflecting that the quantum walker is more localized to the initial state which is not expected in classical scenario.

\subsubsection{Von Neumann Entropy}

Von Neumann entropy measures the degree of mixedness of the quantum state as a result of decoherence~\cite{rieffel2011quantum}. It is defined as:

\begin{equation}
    S(\rho(t)) = -\text{Tr}(\rho(t) \log \rho(t)).
\end{equation}This entropy quantifies the loss of coherence in the quantum state due to interaction with the environment, indicating how much the state has deviated from a pure quantum state towards a mixed state. High entropy indicates a more mixed quantum state due to decoherence, while low entropy reflects a purer quantum state. When there is no noise entropy will stay zero.
\subsubsection{Quantum-Classical Distance}

The quantum-classical distance $D_{QC}(t)$ measures the deviation between the quantum evolution and the classical evolution of the system~\cite{gualtieri2020quantum}. It is defined as:

\begin{equation}
    D_{QC}(t) = 1 - \min_{\rho_{0}} F \left( \rho_{Cl}(t), \rho_{Q}(t) \right),
\end{equation}where $F$ represents the fidelity between the classical and quantum density matrices. The classical density matrix $\rho_{Cl}(t)$ is given by:

\begin{equation}
    \rho_{Cl}(t) = \sum_{k=1}^N p_{k j}(t) |k\rangle \langle k|,
\end{equation}with $p_{k j}(t)$ denoting the classical transition probability from node $j$ to node $k$:

\begin{equation}
    p_{k j}(t) = \langle k | e^{-L t} | j \rangle.
\end{equation}The quantum-classical distance is a measure that tells the divergence between quantum and classical evolution. If it is $0$, then the quantum evolution is fully classicalized, which is not desired, and if $1$, where the system evolves entirely different from classical evolution, indicates that decoherence has not classicalized the system and quantum properties are preserved. Quantum classical distance for complete, cycle and star topologies is already studied in~\cite{bressanini2022decoherence}. 

Using these metrics on different topologies, we analyze which topology is more preferred under different decoherences. 

\section{Results}\label{section4}

\subsection{CTQW on Complete, Cycle and Star Networks}\label{section3a}

\begin{figure}[htbp]
 \begin{tikzpicture}[remember picture, overlay]
 
    \node[rotate=90, anchor=center] at (-0.2, -3.5) {\scriptsize \textbf{No noise}};
    \node[rotate=90, anchor=center] at (-0.2, -7.5) {\scriptsize \textbf{Haken-Strobl}};
    \node[rotate=90, anchor=center] at (-0.2, -11) {\scriptsize \textbf{Intrinsic Decoherence}};
    \node[rotate=90, anchor=center] at (-0.2, -15) {\scriptsize \textbf{Quantum Stochastic}};
\end{tikzpicture}

\begin{center}
    \includegraphics[width=\textwidth]{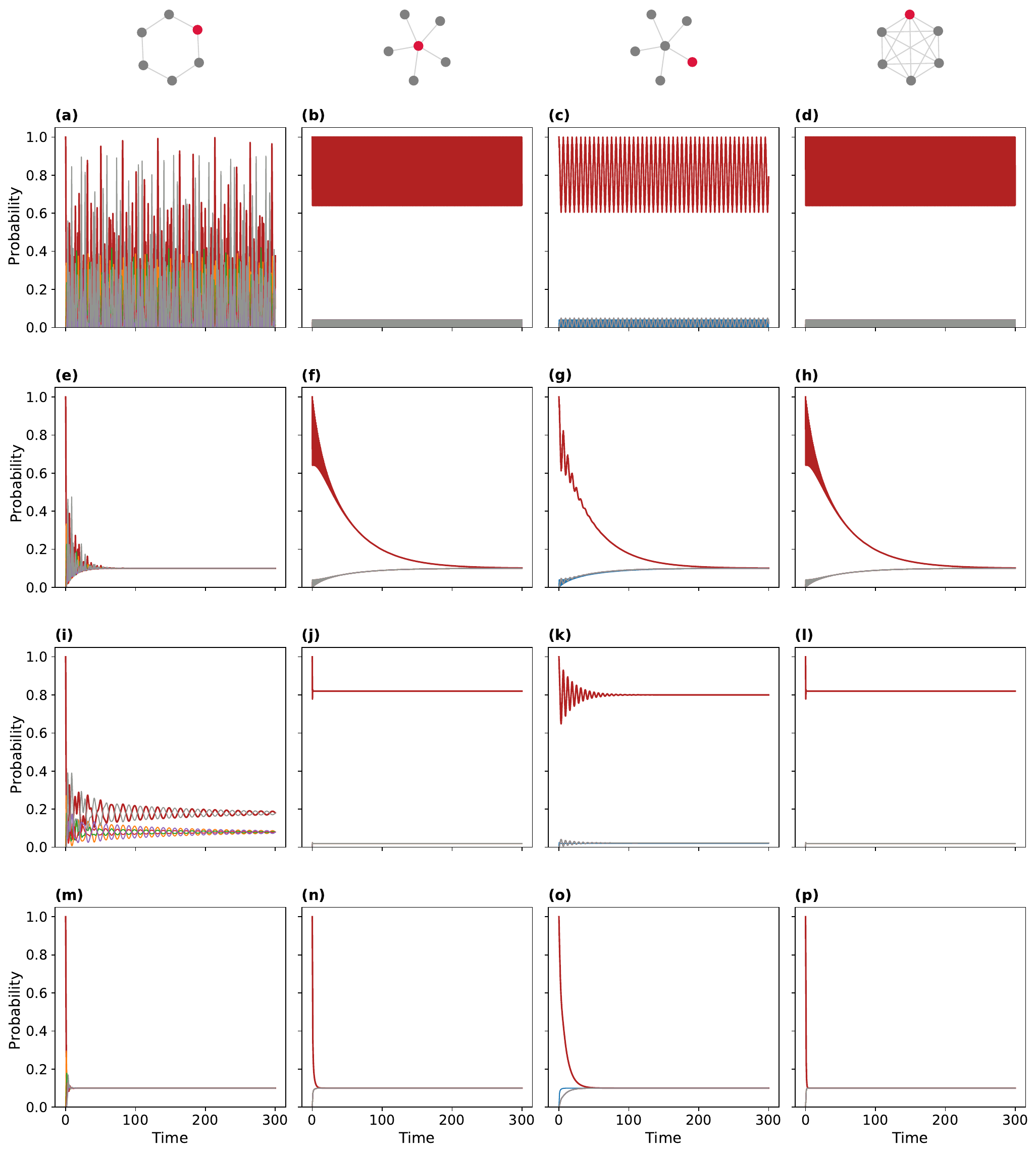}
\end{center}
    \caption{\textbf{Probability distribution of the CTQW on simple network topologies.} To examine the probability distribution of CTQW on various network structures, we consider three representative topologies: cycle, star, and complete graphs, each consisting of $N=10$ nodes. In each network, the red-colored node indicates the site where the walker is initially localized with probability 
    $1$. The temporal evolution of the probability distribution is illustrated under four scenarios: \textcolor{magenta}{(a)–(d)} the noiseless case, \textcolor{magenta}{(e)–(h)} Haken–Strobl noise, \textcolor{magenta}{(i)–(l)} intrinsic decoherence, and \textcolor{magenta}{(m)–(p)} QSW. The columns from left to right correspond to the cycle, star (hub initialization), star (peripheral initialization) and complete network.
    }
    \label{fig:5}
\end{figure}
\begin{figure}[htbp]
    \centering
    \includegraphics[width=1\textwidth]{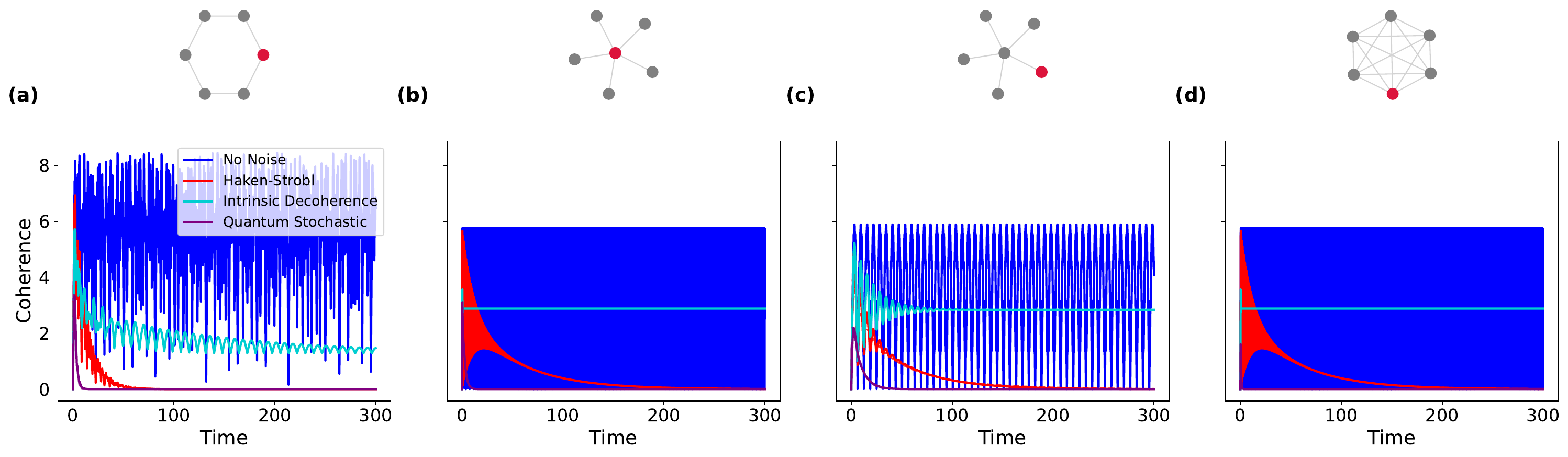}
    \includegraphics[width=\textwidth]{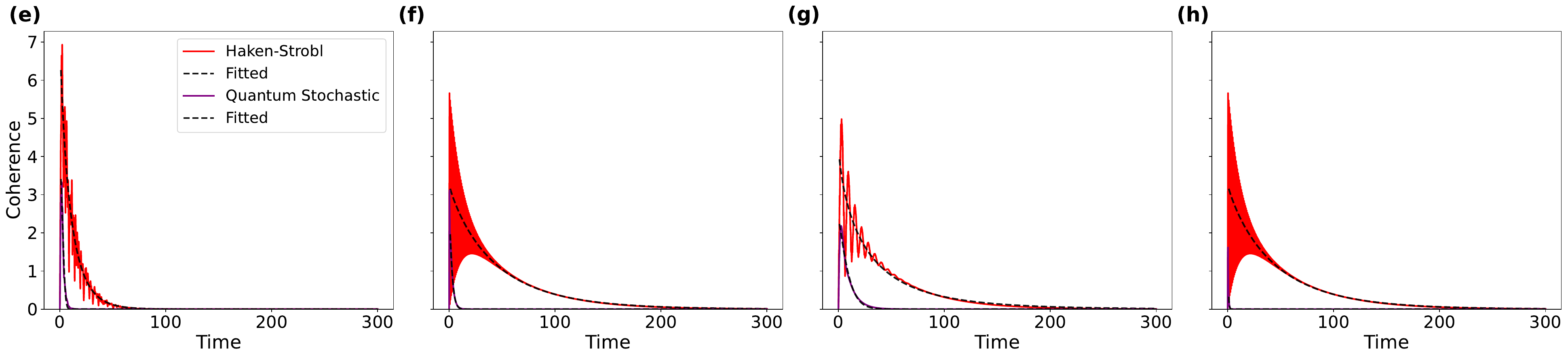}
    \caption{\textbf{$\ell_1$-norm of coherence with time for simple topologies with network size N = 10.} In the first row, $\ell_1$-norm of coherence is plotted for all types of decoherence for cycle \textcolor{magenta}{(a)}, star network with hub node initialization \textcolor{magenta}{(b)} and peripheral node initialization \textcolor{magenta}{(c)}, and complete network \textcolor{magenta}{(d)}. The second row presents \textcolor{magenta}{(e)-(f)} the coherence decay profiles for both Haken–Strobl noise and QSW, with the data fitted using the proposed decay model in Eq.~(\ref{decay_fit}), with the best-fit parameters summarized in Table~\ref{tab:coherence_decay}.
    }
    \label{fig:6}
\end{figure}

To investigate the CTQW dynamics on different network topologies, we first analyze the probability distribution of the CTQW on each node for the cycle, complete, and star networks. For the star network, we consider two cases; in one case, the walker is initialized at the hub, and in another case, it is initialized at any peripheral node. For the cycle and complete networks, owing to degree symmetry, it starts from a randomly selected node. In all cases, the initially localized node has probability $1$ at $t=0$. To assess system stability, we additionally compute the $\ell_1$-norm of coherence, the quantum–classical distance, the fidelity with the initial state, and the von Neumann entropy. For all analyzes, we fix the network size at $N=10$, as the $\ell_1$-norm of coherence for star networks of different sizes (with hub initialization) exhibits qualitatively similar behavior; second, large-sized networks preserve coherence longer under Haken-Strobl and intrinsic decoherence (see Fig.~\ref{fig:coherence_p_gamma}\textcolor{magenta}{(a)-(c)} in Appendix~\ref{scaling for simple}). The decoherence parameters are set to $\gamma=0.1$ and $p=0.1$, since these values are sufficiently large to make decoherence effects clearly visible (Fig.~\ref{fig:coherence_p_gamma}\textcolor{magenta}{(d)-(f)}).

Temporal probability plots (Fig.~\ref{fig:5}) show that, in the absence of noise, the CTQW on the cycle network (Fig.~\ref{fig:5}\textcolor{magenta}{(a)}) explores not only the initially localized node but also other nodes in the network. In contrast, for both the star and complete networks (Fig.~\ref{fig:5}\textcolor{magenta}{(b)-(d)}), the walker remains strongly localized at the initial node, maintaining a high occupation probability. The identical behavior of the hub-initialized star and complete networks, owing to the fully connected ($N-1$) initial node is consistent with the universality reported by Razzoli \textit{et al.}~\cite{razzoli2022universality}, who showed that, in the noiseless case, CTQW dynamics initiated at a fully connected hub depend only on the network size and not on the remaining graph structure. Notably, our results indicate that this universality is not only true for noiseless case, but true for Haken-Strobl and intrinsic decoherence as shown in Fig.~\ref{fig:5}\textcolor{magenta}{(f),(h),(j),(l)}.

We next analyze the temporal behavior of the $\ell_1$-norm of coherence (Fig.~\ref{fig:6}). We observe that the cycle network exhibits persistent oscillations that do not decay to zero (Fig.~\ref{fig:6}\textcolor{magenta}{(a)}) and has the highest coherence among all topologies. This reflect that network topologies that don't have hub nodes promotes delocalized dynamics and enhanced coherence, thereby reflecting a trade-off between localization and coherence in the noiseless case. In contrast, the complete network and the hub-initialized star network (Fig.~\ref{fig:6}\textcolor{magenta}{(b),(d)}) display higher-frequency oscillations in both occupation probabilities and coherence than the peripheral-initialized star network (Fig.~\ref{fig:6}\textcolor{magenta}{(c)}), consistent with the high-frequency revivals in transition probabilities reported by Xu \textit{et al.}~\cite{xu2009exact}. These results highlight the strong dependence of CTQW dynamics on network structure and initial conditions.

Under Haken-Strobl noise (Fig.~\ref{fig:5}\textcolor{magenta}{(e)-(h)}), all networks converge to a uniform probability distribution, with the cycle network reaching the steady state fastest. This behavior resembles the equilibration of probabilities observed in classical continuous-time random walks~\cite{mulken2011continuous}.
The simultaneous decay of coherence to zero (Fig.~\ref{fig:6}) and the equilibration of probabilities shows the onset of maximally mixed state $\frac{\mathbb{I}}{N}$ and is consistent with the fact that both the Haken-Strobl noise and QSW Liouvillians are unital and Davies irreducible~\cite{davies1976quantum} for connected graphs.  As shown in Appendix~\ref{HSproof} and Appendix~\ref{QSWproof}, irreducibility of the Liouvillians is verified using the algebraic criteria of Zhang and Barthel~\cite{zhang2024criteria}, ensuring a unique maximally mixed steady state.

Under intrinsic decoherence (Fig.~\ref{fig:5}\textcolor{magenta}{(i)-(l)}), the star and complete networks remain strongly localized at the initial node and rapidly approach a steady state. In contrast, the cycle topology does not exhibit strong localization; instead, node occupation probabilities display damped oscillations, and the $\ell_1$-norm of coherence shows persistent temporal fluctuations even at long times (Figs.~\ref{fig:5}\textcolor{magenta}{(i)} and~\ref{fig:6}\textcolor{magenta}{(a)}).  Although coherence survives across all topologies (Fig.~\ref{fig:6}\textcolor{magenta}{(a)-(d)}), it is highest for the star and complete networks than cycle network. For peripheral initialization in the star network (Figs.~\ref{fig:5}\textcolor{magenta}{(k)} and \ref{fig:6}\textcolor{magenta}{(c)}), coherence and initially localized node probability exhibits damped oscillations and a slower relaxation compared to hub initialization, while ultimately converging to the same steady-state value as for complete network.

Under Haken-Strobl and QSW decoherence, the coherence decays with time. To quantify the decay rate of the $\ell_1$-norm of coherence, we fit the simulated data using the stretched-exponential Kohlrausch function~\cite{kohlrausch1854theorie}:

\begin{equation}
C(t) = C_0 \exp\left( - (\lambda t)^\beta \right).
\label{decay_fit}
\end{equation}

Here, $\lambda$ is the effective decay rate that determines how quickly the function decreases over time, while $\beta$ is the stretching exponent that characterizes the nature of the decay.  However, the overall behavior of the coherence decay is determined by the combined influence of both $\lambda$ and $\beta$. This can be quantified as a single characteristic relaxation time, $\tau$, which represents the mean lifetime of the coherence from its normalized initial value~\cite{gueguen2015relationship}:
\begin{equation}
      \tau = \frac{1}{\beta \lambda} \, \Gamma\!\left(\frac{1}{\beta}\right)
      \label{eq:tau}
\end{equation}

 We fitted coherence data obtained for the Haken-Strobl noise and QSW as shown in Fig.~\ref{fig:6}\textcolor{magenta}{(e)-(h)} and the extracted coefficients $\lambda$, $\beta$ and $\tau$ summarized in Table~\ref{tab:coherence_decay}. The complete and hub initialized star networks exhibit slower coherence decay characterized by the highest $\tau$, followed by the star peripheral initialized case, with the cycle network showing the fastest decay in the case of Haken-Strobl noise. We also observe that coherence decays significantly faster under QSW decoherence than under Haken–Strobl for all network topologies, and the stability ranking correspondingly inverted. Specifically, under QSW, the star graph with peripheral node initialization take longer to equalize probability than the star with hub initialization and the complete network, whereas the opposite trend is observed under Haken–Strobl decoherence. Notably, the peripheral node initialized star and cycle network exhibits comparatively slower coherence decay under QSW, indicating enhanced robustness. Since QSW is a link-based decoherence model (Eq.~\eqref{eq.QSW}), dissipation increases with the number of connections, rendering highly connected structures such as the complete graph and star-center configurations more fragile, while minimally connected configurations, such as peripheral-node initialization, remain more stable. To further elucidate the origin of these stability hierarchies, Appendix~\ref{sec.network_properies_and_tau} establishes a direct correlation between the relaxation time $\tau$ and specific properties of the initialized node. These results suggest that the closeness centrality is important to capture actual $\tau$ for the simple networks under Haken-Strobl decoherence and inverse density weighted degree for the QSW decoherence. 
 Overall our result suggest that in single CTQW, the preservation of coherence and stability depends on the network structure, choice of initial localization and decoherence types. 
 
\begin{table}[htbp]
    \centering
    \begin{tabular}{|c|l|c|c|c|}
        \hline
        \textbf{Decoherence} & \textbf{Network} & $\lambda$ & $\beta$& $\tau$\\
        \hline
        Haken-Strobl & Cycle         & 0.14 & 0.86 & 7.72 \\
                     & Star Center  & 0.03 & 0.90 & 35.07 \\
                     & Star peripheral     & 0.10 & 0.71&12.49 \\
                     & Complete      & 0.03 & 0.90& 35.07  \\
        \hline
        Quantum Stochastic& Cycle         & 0.08  & 2.03&11.08\\
                                & Star Center   & 0.28  & 2.62&3.17 \\
                                & Star peripheral     & 0.06  & 1.23 &15.58 \\
                                & Complete      & 2.00  & 1.00&0.50\\
        \hline
    \end{tabular}
    \caption{Coherence decay fitting parameters for cycle, star, and complete networks under Haken-Strobl and QSW decoherence.}
    \label{tab:coherence_decay}
\end{table}

Further, the stability of CTQW is systematically analyzed in Appendix~\ref{sec.complete_cycle} through quantum-classical distance, fidelity with the initial state (Fig.~\ref{fig:qc_fid_combined_2}), and von Neumann entropy (Fig.~\ref{fig:14_2}). Our result show that
under Haken-Strobl noise and intrinsic decoherence, the star network (with hub node initialization) and the complete graph exhibit high stability, indicated by elevated quantum-classical distance and fidelity, along with low entropy for Haken-Strobl noise. For the QSW, the star network initialized at a peripheral node shows improved stability, as seen from its high fidelity, low entropy and higher value of quantum-classical distance.

In summary, star network with peripheral initialization shows greater stability under QSW decoherence. However, for all other decoherence models, the complete graph and the star network with central node initialization outperform other topologies offering enhanced stability. Furthermore, the cycle network exhibits resistance to intrinsic decoherence induced steady state even at long times.

\subsection{CTQW on Complex Network Topologies}\label{section3b}
We extend our analysis to more complex topologies to investigate how different complex network structures influence the stability of quantum systems. Specifically, we study the evolution of CTQW on Erdős–Rényi, small-world, and scale-free networks, maintaining a fixed average degree of $\langle k \rangle = 4$ and network size $N=100$; the broader impact of varying average connectivity and network size is detailed in Appendix~\ref{averagre} (Fig.~\ref{Fig:average_degree} and Fig.~\ref{Fig:complex_size}). As the degree of the initially localized node can also significantly affect the dynamics (see Appendix~\ref{initial_scalefree} Fig.~\ref{fig:10}), we choose a high-degree node as the initial localization in the heterogeneous scale-free network. In contrast, the Erdős–Rényi and small-world networks have nodes with approximately similar average degrees, so the impact of the initial node selection is expected to be comparatively minimal.

\begin{figure*}[htbp]
    \centering
    \begin{tikzpicture}

    \node[rotate=90, anchor=center] at (-0.75,8.69)
        {\footnotesize{Erdős--Rényi}};
    \node[rotate=90, anchor=center] at (-0.75, 4.8)
        {\footnotesize{Small--World}};
    \node[rotate=90, anchor=center] at (-0.75, 0.9)
        {\footnotesize{Scale--Free}};

    \node[anchor=south west, inner sep=0] at (0,7)
        {\includegraphics[width=1.01\textwidth]{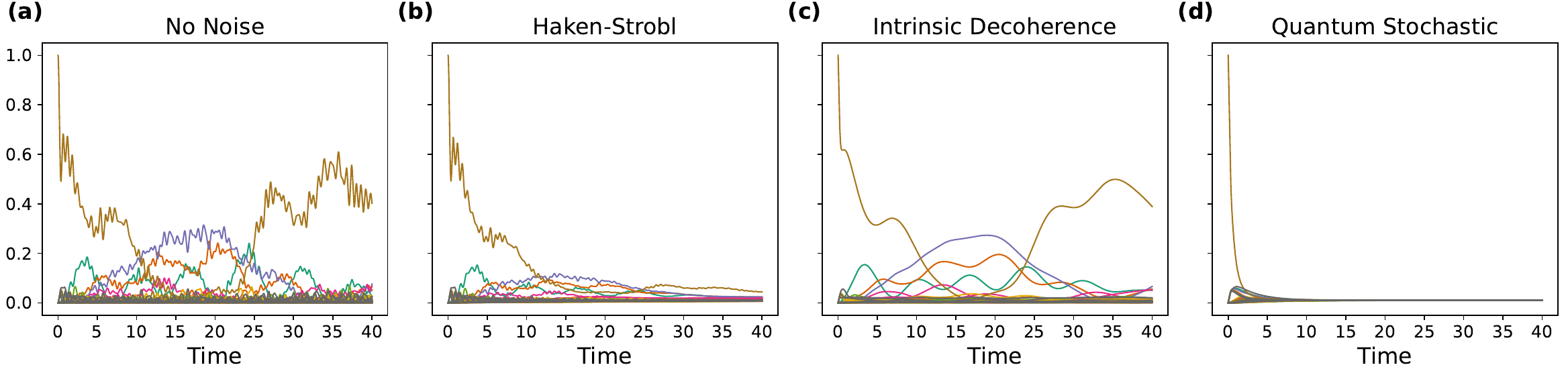}};
    \node[rotate=90, anchor=center, opacity=0.75] at (-0.2,8.7)
        {\scriptsize Probability};

    \node[anchor=south west, inner sep=0] at (0,3)
        {\includegraphics[width=1.01\textwidth]{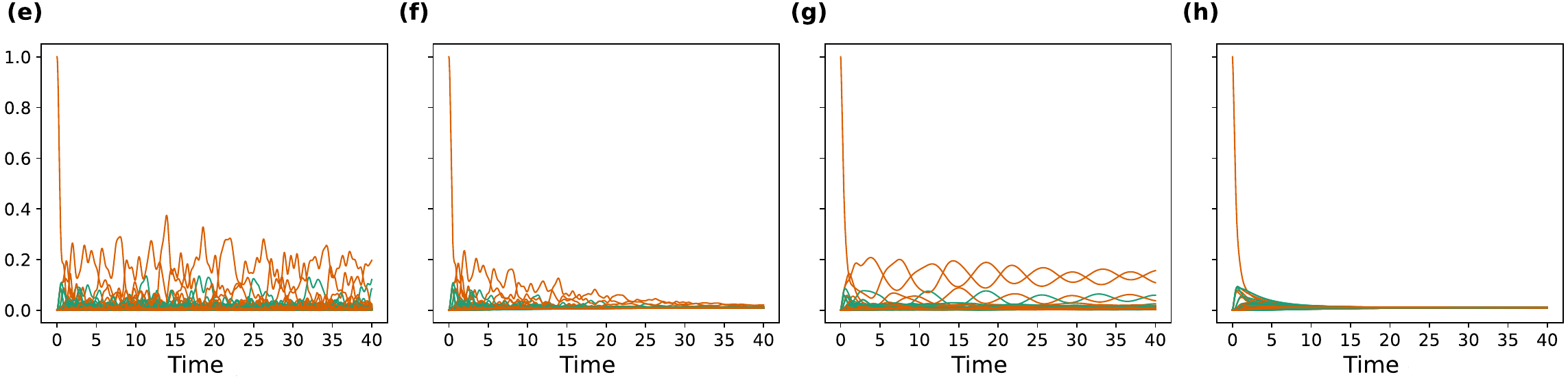}};
    \node[rotate=90, anchor=center, opacity=0.75] at (-0.2,4.8)
        {\scriptsize Probability};

    \node[anchor=south west, inner sep=0] at (0,-1)
        {\includegraphics[width=\textwidth]{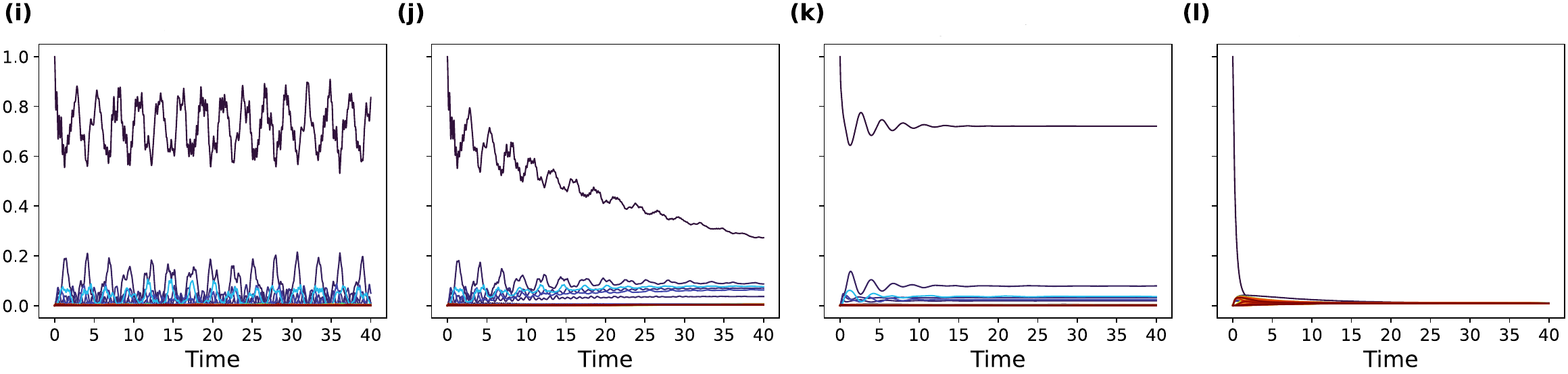}};
    \node[rotate=90, anchor=center, opacity=0.75] at (-0.2,0.9)
        {\scriptsize Probability};

    \end{tikzpicture}
    \caption{\textbf{Probability versus time plots on complex network topologies with network size N = 100 under various decoherence models}. The top row shows the probability distribution of the CTQW on an Erdős–Rényi network, with the walker initially localized on a randomly selected node at time $t=0$ \textcolor{magenta}{(a)–(d)}. The middle row depicts the dynamics on a small-world network \textcolor{magenta}{(e)–(h)}, while the bottom row presents the probability evolution on a scale-free network, where the walker is initially placed on a high-degree node \textcolor{magenta}{(i)-(l)}. For each network, the probability distribution across all nodes is shown over time, with the occupation probability of each node represented using a distinct color.}
    \label{fig:11}
\end{figure*}

\begin{figure*}[htbp]
  \centering
    \begin{minipage}{0.32\textwidth}
        \centering
        \footnotesize{Erdős–Rényi Network} \\[0.5ex]
        \includegraphics[width=\textwidth]{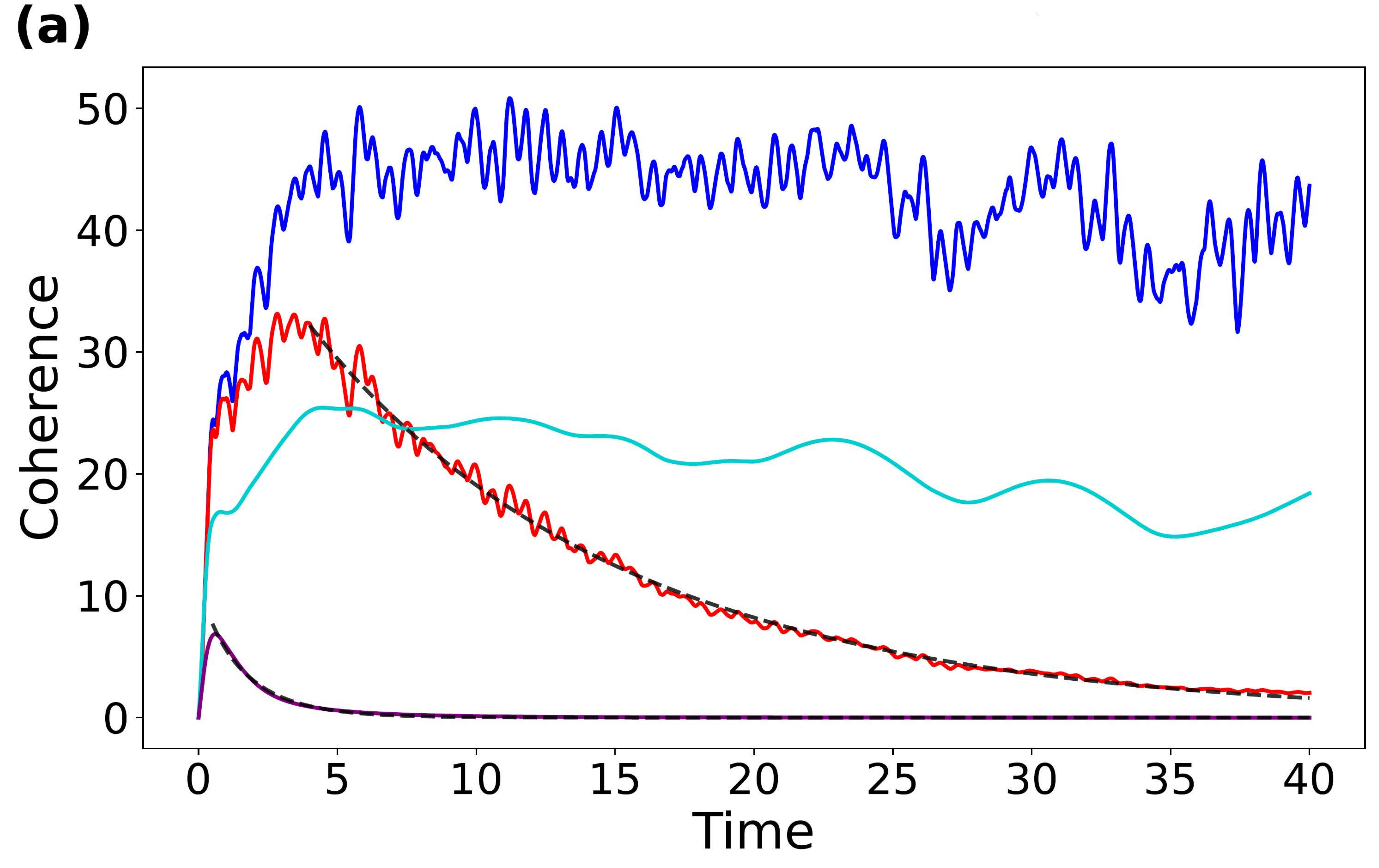}
    \end{minipage}
    \hfill
    \begin{minipage}{0.32\textwidth}
        \centering
        \footnotesize{Small-World Network} \\[0.5ex]
        \includegraphics[width=\textwidth]{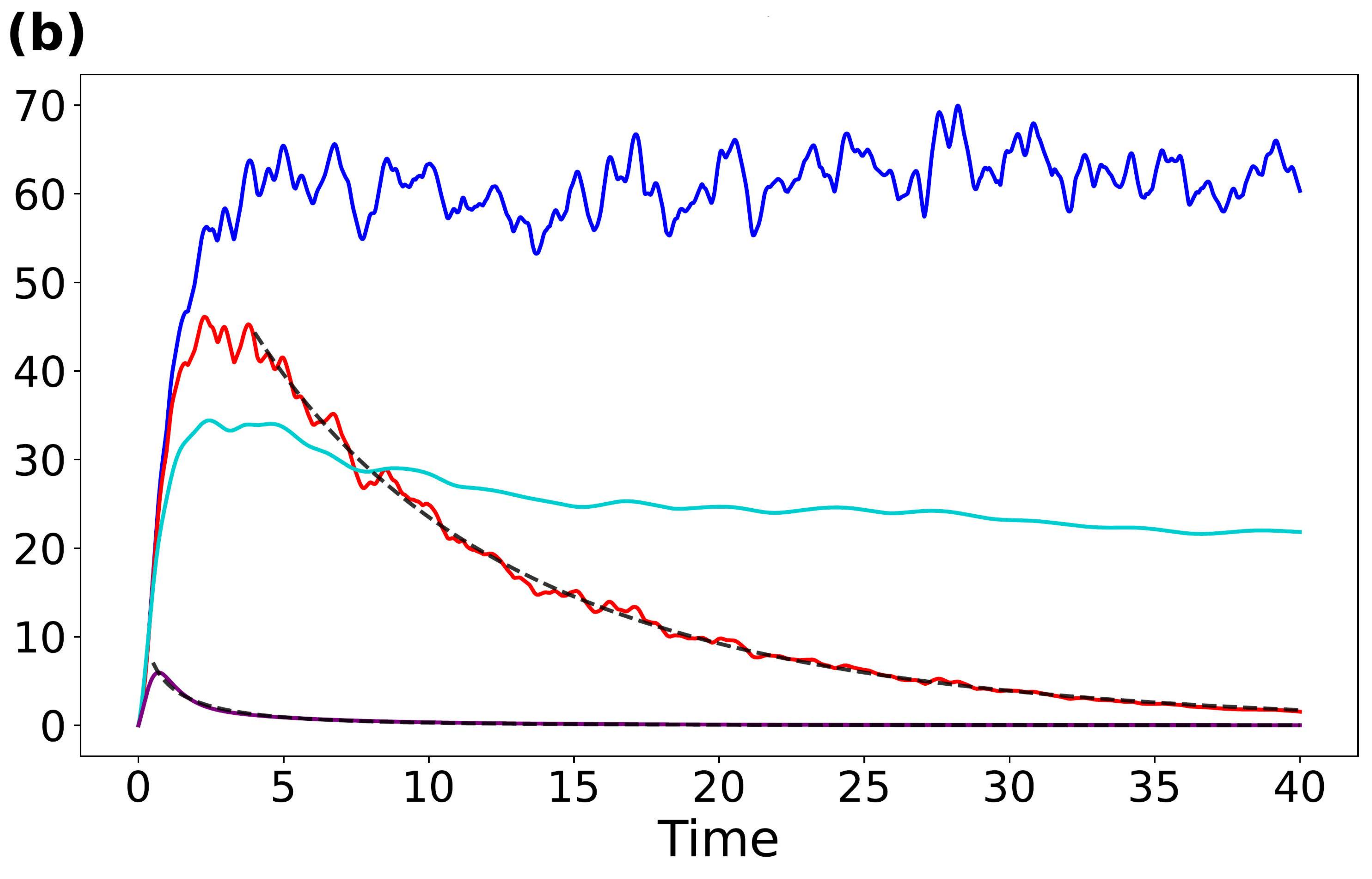}
    \end{minipage}
    \hfill
    \begin{minipage}{0.32\textwidth}
        \centering
        \footnotesize{Scale-Free Network} \\[0.5ex]
        \includegraphics[width=\textwidth]{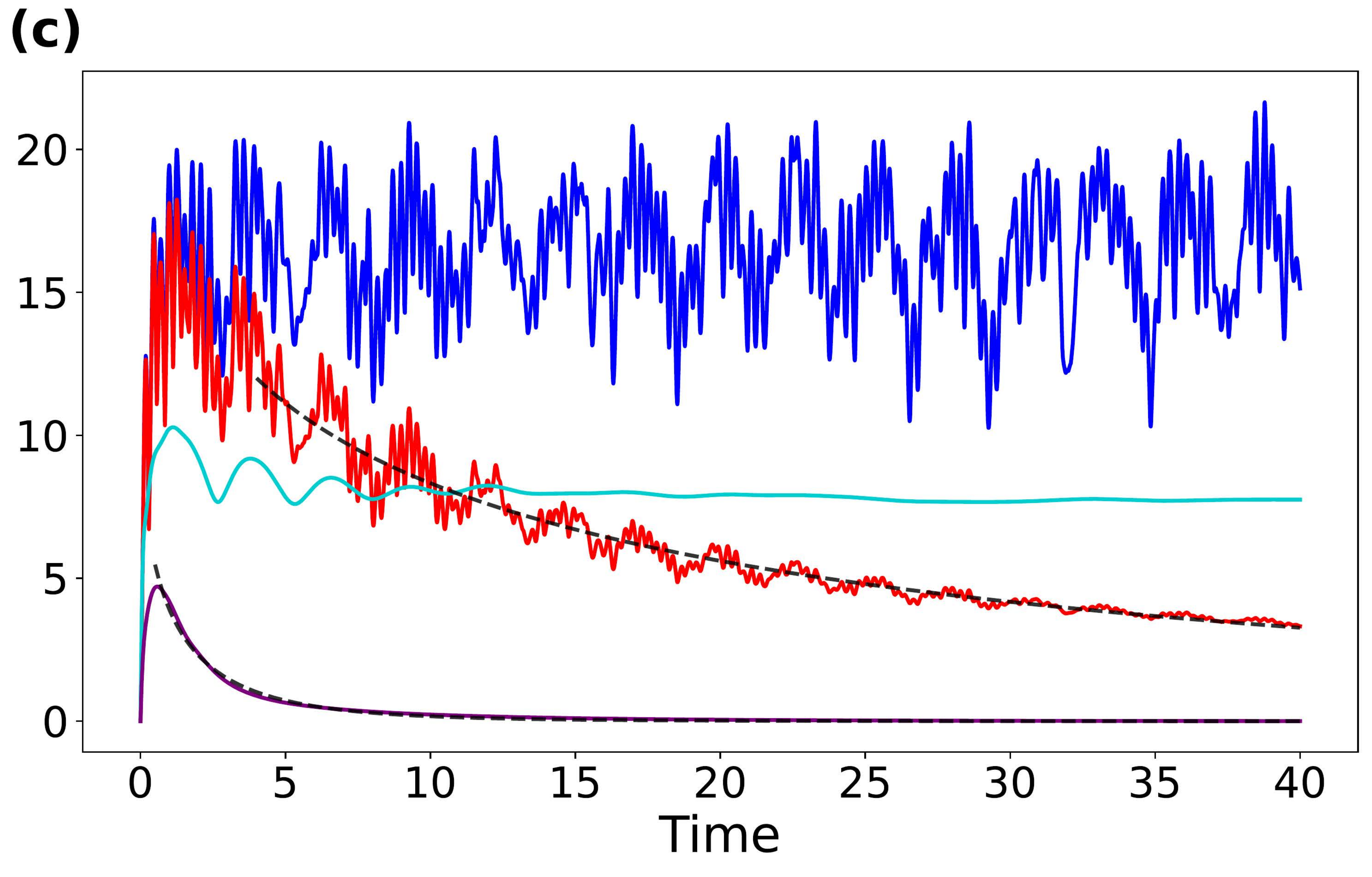}
    \end{minipage}

    \caption{\textbf{Time evolution of the $\ell_1$-norm of coherence for complex network topologies with size N = 100.} The coherence decay under Haken–Strobl noise and QSW is modeled using the decay function given in Eq.~(\ref{decay_fit}), with the best-fit parameters summarized in Table~\ref{tab:network_decay}. The $\ell_1$-norm of coherence is plotted for the Erdős–Rényi network \textcolor{magenta}{(a)}, small-world network \textcolor{magenta}{(b)}, and scale-free network \textcolor{magenta}{(c)} under four conditions: no noise (blue), Haken–Strobl noise (red), intrinsic decoherence (cyan), and quantum stochastic walk (purple). These curves illustrate how coherence diminishes or preserve over time depending on the network structure and the specific decoherence model.}
    \label{fig:12}
\end{figure*}

 Fig.~\ref{fig:11} shows strong localization at the initially occupied hub node in the scale-free network. In contrast, Erdős–Rényi and small-world networks lack pronounced degree heterogeneity and central hubs, leading to a more delocalized evolution of the CTQW. 
 Scale-free networks are heterogeneous, so the evolution of CTQW also strongly depends on the centrality of the node where the walker is initially localized (see Appendix~\ref{initial_scalefree} Fig.~\ref{fig:10}). 
Depending on the network configuration, a given node may or may not simultaneously possess multiple centrality maxima. For example, a node may have both the highest degree and the highest closeness centrality. However, when these maxima occur at different nodes and the walker is initialized at one of them, the occupation probabilities of walker at the highest-degree and highest-closeness nodes exhibit an anti-phase synchronization pattern. When the initialized node has both the highest degree and closeness centrality, the dynamics exhibit strong localization at that node, except in the presence of QSW decoherence. Xu \textit{et al.}~\cite{xu2009exact} demonstrated that, in the thermodynamic limit ($N \to \infty$), transition probabilities in star networks remain localized at the initial site in noiseless case. Since both star and scale-free networks are heterogeneous and contain hub nodes, our results suggest that localization at a centrally initialized node is a general feature of such networks, valid not only in the noiseless case but also under intrinsic decoherence. Further, Fig.~\ref{fig:12} shows that, in the absence of decoherence, both small-world and Erdős–Rényi networks maintain a higher $\ell_1$-norm of coherence compared to the scale-free network. This highlights the influence of network structure, suggesting that while hubs facilitate localization, they may simultaneously contribute to reduced coherence in noiseless conditions.

Under the influence of Haken–Strobl noise, the occupation probability at the initially localized node decays rapidly in both the Erdős–Rényi and small-world networks Fig.~\ref{fig:11}\textcolor{magenta}{(b),(f)}, eventually leading to a uniform distribution. This renders the initially localized node indistinguishable from others, indicating complete delocalization across the network. This process is accompanied by the $\ell_1$-norm of coherence decaying to zero (Fig.~\ref{fig:12}\textcolor{magenta}{(a),(b)}), making the transition to a steady state (Appendix~\ref{HSproof}). In contrast, the scale-free network does not exhibit complete delocalization in short time Fig.~\ref{fig:11}\textcolor{magenta}{(j)}. Although the occupation probability at the initially localized node decays over time, the distribution remains non-uniform and approaches uniformity much more slowly than in the Erdős-Rényi and small-world networks. The scale-free network  shows enhanced robustness under Haken-Strobl noise, preserving coherence for longer times than the other topologies (Fig.~\ref{fig:12}\textcolor{magenta}{(c)}). In contrast, under QSW decoherence it exhibits the opposite trend. All topologies delocalize and lose coherence significantly faster under QSW decoherence than under the other noise models, as evident from Figs.~\ref{fig:11} and \ref{fig:12}.

\begin{table}[htbp]
    \centering
 \begin{tabular}{|c|l|c|c|c|c|}
        \hline
        {\textbf{Decoherence}} & {\textbf{Network}} & \multicolumn{3}{c|}{\textbf{Single Realization}} & \textbf{Ensemble} \\
        \cline{3-6}
        & & $\lambda$ & $\beta$ & $\tau$ & $\langle \tau \rangle$ \\
        \hline
        {Haken-Strobl} & Erdős-Rényi  & 0.10 & 0.84 & 10.96 & 11.46 \\
                                     & Small-World  & 0.17 & 0.96 & 5.99  & 11.56 \\
                                     & Scale-Free   & 0.24 & 0.37 & 17.43 & 38.99 \\
        \hline
        {\begin{tabular}[c]{@{}c@{}}Quantum Stochastic\end{tabular}} & Erdős-Rényi  & 0.71 & 0.45 & 3.49  & 2.17 \\
                                     & Small-World  & 1.53 & 0.90 & 0.69  & 3.46 \\
                                     & Scale-Free   & 0.96 & 0.62 & 1.50  & 1.72 \\
        \hline
    \end{tabular}
    \caption{
    Coherence decay fitting parameters and relaxation time are for Erdős-Rényi, small-world, and scale-free topologies. Values include for both individual realizations ($\tau$) and ensemble-averaged relaxation time ($\langle \tau \rangle$)  calculated for 50 realizations under Haken-Strobl and QSW decoherence models.}
    \label{tab:network_decay}
\end{table}

Further, the decay coefficients $\lambda$ and $\beta$ are obtained by fitting the $\ell_1$-norm of coherence (Fig.~\ref{fig:12}), from which we compute the relaxation time $\tau$ (Table~\ref{tab:network_decay}) and the ensemble-averaged relaxation time $\langle \tau \rangle$ over $50$ network realizations. These quantitative measures verify our earlier observations that under Haken-Strobl noise, the scale-free networks preserve coherence for longer times than the QSW, as reflected by its comparatively larger values of $\tau$ and $\langle \tau \rangle$.

 Conversely, QSW decoherence induces the most rapid decay across all complex 
 topologies, with the scale-free network proving more vulnerable than Erdős–Rényi as shown by a lower value of $\langle \tau \rangle$. Such relative vulnerability is consistent with the findings for simple topologies discussed in Sec.~\ref{section3a}, and stems from the fundamental nature of the type of decoherence mechanisms. Specifically, since QSW jumps are link-dependent, the high degree of the hub nodes in heterogeneous networks causes a higher decoherence and consequently faster decay of coherence. To systematically bridge the network's structural features with the observed dynamics, Appendix~\ref{sec.network_properies_and_tau} explores $\langle \tau \rangle$, and the statistical average of initialized node properties. Our results suggest that different properties of initially localized node contribute to capture the actual dynamics of the quantum system under different decoherence models.

In the presence of intrinsic decoherence, the CTQW exhibits partial localization at the initially localized node in the scale-free network, reaching decoherence-induced steady state Fig.~\ref{fig:11}\textcolor{magenta}{(k)}. In contrast, localization behavior of CTQW is not observed in the Erdős–Rényi and small-world networks, which also resist reaching a steady-state within the observed time-scales as illustrated in Fig.~\ref{fig:11}\textcolor{magenta}{(c),(g)}.  
Interestingly, the $\ell_1$-norm of coherence remains preserved under intrinsic decoherence for all network topologies (Fig.~\ref{fig:12}), consistent with the behavior observed for simple networks in Sec.~\ref{section3a}. This indicates that, although intrinsic decoherence induces partial localization in the scale-free network, it preserves a finite amount of quantum coherence across all topologies, thereby maintaining stability.

\begin{figure*}[htbp]
    \centering

        \centering
        \includegraphics[width=\textwidth]{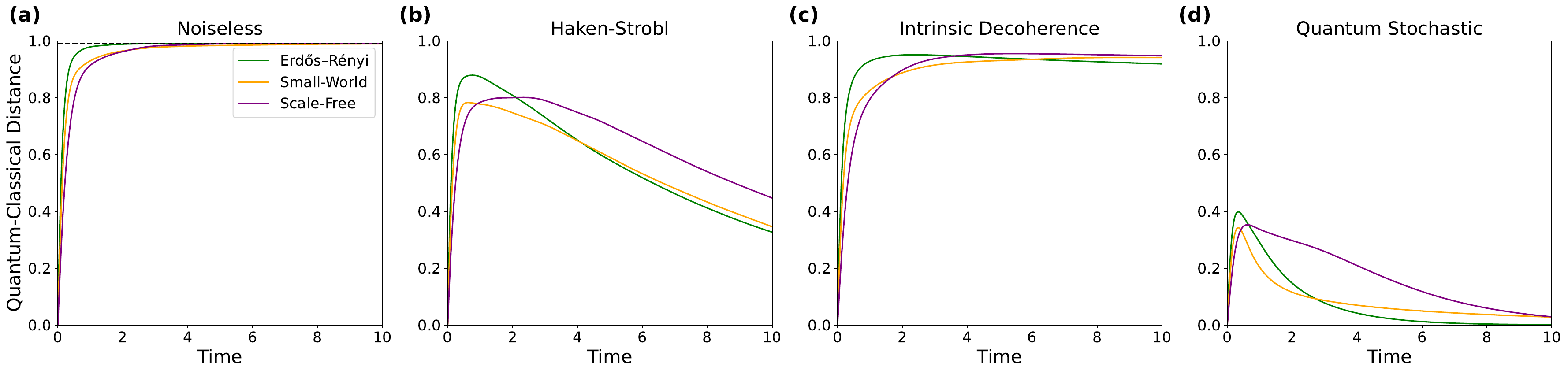}

        \centering
        \includegraphics[width=\textwidth]{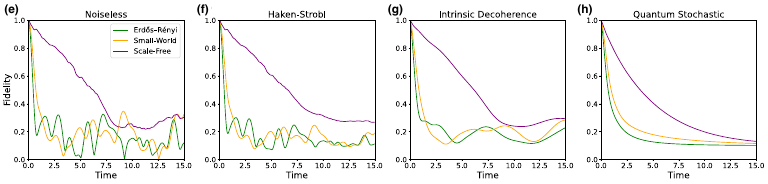}

    \caption{\textbf{Quantum-classical distance and fidelity with the initial state for complex network topologies with size N = 100}. Quantum-classical distance and fidelity with the initial state for Erdős-Rényi (green), small-world (yellow), and scale-free (magenta) networks under the absence of noise \textcolor{magenta}{(a),(e)} and three decoherence models: Haken-Strobl \textcolor{magenta}{(b),(f)}, intrinsic Decoherence \textcolor{magenta}{(c),(g)}, and QSW \textcolor{magenta}{(d),(h)}.}
    \label{fig:qc_fid_combined}
\end{figure*}

\begin{figure*}[htbp]
    \centering
    \includegraphics[width=0.75\textwidth]{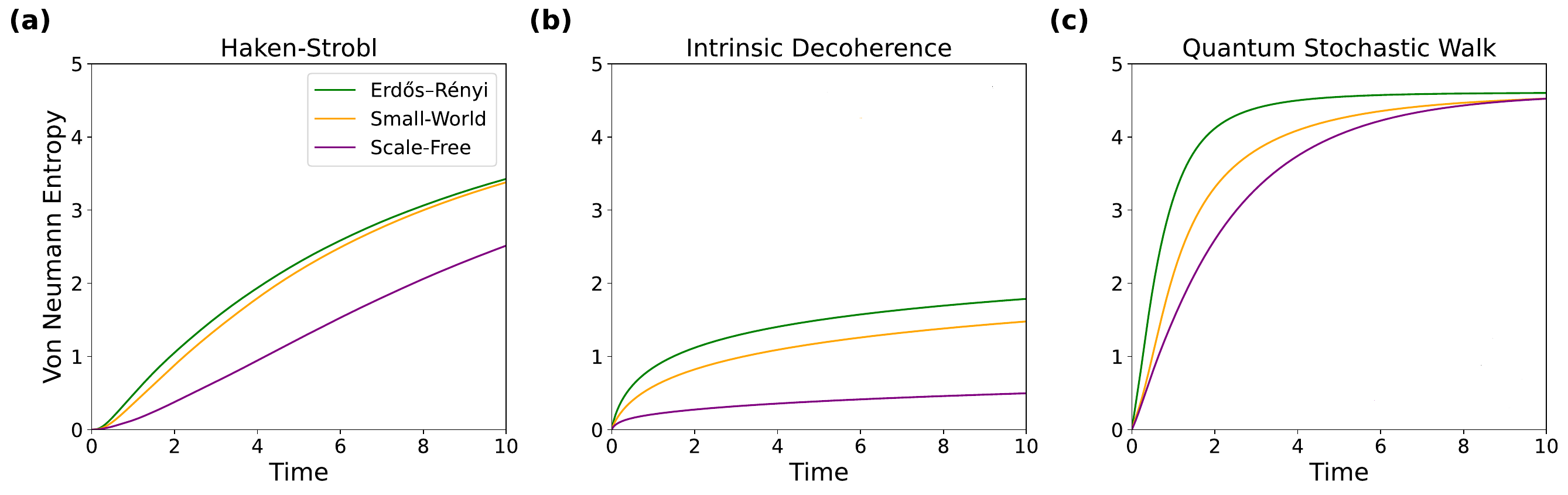}
    \caption{\textbf{Time evolution of von Neumann entropy for complex network topologies for size N = 100}. Von Neumann entropy for Erdős-Rényi (green), small-world (yellow), and scale-free (magenta) networks under Haken-Strobl \textcolor{magenta}{(a)}, Intrinsic Decoherence \textcolor{magenta}{(b)}, and QSW \textcolor{magenta}{(c)}.}
    \label{fig:14}
\end{figure*}

Further we use other standard metrics namely quantum-classical distance, von Neumann entropy, and fidelity to check the stability of quantum system. 
The quantum-classical distance in Fig.~\ref{fig:qc_fid_combined}\textcolor{magenta}{(a)-(d)} suggest that the scale-free network exhibits greater resistance to classicalization under decoherence compared to the Erdős–Rényi and small world network, since as time progresses, scale-free has a higher quantum classical distance than other topologies underlying all decoherence dynamics. 
It shows a slower decay of quantum-classical distance in the presence of both QSW and Haken-strobl both of which classicalize the system over long timescales. For intrinsic decoherence in Fig.~\ref{fig:qc_fid_combined}\textcolor{magenta}{(c)}, the distance does not decay to zero, indicating that classicalization does not occur and quantum features are consistently preserved for all networks. For the noiseless scenario, across all network topologies, the quantum-classical distance converges to the theoretical value $\lim_{t \to \infty} D_{QC}(t) = 1 - \frac{1}{N}$~\cite{gualtieri2020quantum}, as represented by the dashed line in Fig.~\ref{fig:qc_fid_combined}\textcolor{magenta}{(a)}.

The fidelity plots in Fig.~\ref{fig:qc_fid_combined}\textcolor{magenta}{(e)-(h)} show that the scale-free network topology exhibits localization near the initial state since it has high fidelity with the initial state. The higher the fidelity lesser the evolved state has deviated from the initial state. On contrast, the Erdős-Rényi and small-world networks have a little less fidelity across various decoherence. 

Von Neumann entropy plots Fig.~\ref{fig:14}\textcolor{magenta}{(a)-(c)} show that entropy is high for Erdős-Rényi in all decoherence. Entropy is least for scale-free in the case of Haken-Strobl and QSW, showing scale-free is not much deviated from the pure state as compared to Erdős-Rényi.  This implies that scale-free is more resilient to all noises and Erdős-Rényi is more prone to mixed states caused by decoherence than other topologies. For the noiseless case, the entropy always remains zero. 

Taken together, despite the rapid coherence decay under QSW decoherence, scale-free networks consistently maintain higher quantum-classical distance, higher fidelity, and lower entropy, demonstrating their superior stability relative to other topologies.

\section{Conclusion and Discussion}\label{section5}
In this work, we investigate the stability of CTQW by analyzing its dynamics on various network topologies under different decoherences. We define stability in terms of the preservation of quantum properties over time and examine how this preservation is influenced by the interplay between network structure and decoherence model. 

To quantify the stability, we first compute the node probability distribution of the CTQW and the $\ell_1$-norm of coherence. These measures confirm that, under Haken-Strobl and QSW decoherence, the system Liouvillians are unital and Davies-irreducible on connected networks (~\ref{proof}), and therefore relax to the maximally mixed steady state. This contrasts with the intrinsic decoherence model, which allows for non-trivial steady states and localization. 

Further, our results suggest that stability varies significantly across decoherence models, intrinsic decoherence preserves coherence most effectively, followed by Haken–Strobl noise, while QSW results in the most rapid coherence loss.
Our results reveal that coherence preservation depends not only on the type of decoherence but also on the underlying topology and the specific node at which the quantum walker is initialized, specifically in heterogeneous networks. 

Delving deeper, we find that sparse and relatively homogeneous networks, such as the cycle, Erdős-Rényi, and small-world topologies, exhibit stronger probability delocalization. In contrast, dense and heterogeneous networks, including the complete, star, and scale-free topologies, show pronounced localization at the initially occupied node in the noiseless and intrinsic decoherence cases, while simultaneously exhibiting higher coherence values. 

Under Haken-Strobl noise, the hub-initialized star, scale-free, and complete networks retain coherence longer than the other topologies. However, in the QSW model, the peripheral-initialized star network, along with Erdős--Rényi and small-world networks, demonstrates comparatively longer coherence preservation. These trends are clearly reflected in both the probability distributions and the $\ell_1$-norm of coherence. Quantitative analysis based on fitting the decay of the $\ell_1$-norm of coherence further supports these observations. In particular, extracting the relaxation time from coherence decay provides a direct measure of decoherence loss and establishes a connection between stability timescales and underlying network structural properties.

In contrast, other stability quantifiers do not always provide a uniform characterization. For instance, fidelity and von Neumann entropy often indicate the highest stability for scale-free networks across all decoherence models. Similarly, the quantum-classical distance suggests enhanced stability of scale-free networks under Haken-Strobl and QSW dynamics. However, in the noiseless and intrinsic decoherence regimes, the quantum-classical distance yields nearly identical trajectories for all complex networks, making it difficult to clearly distinguish their relative stability.

Our results also support the previous studies that the interpretation of stability is inherently application dependent. For example, high entropy may impair the efficiency of quantum search algorithms~\cite{bose2000communication}, while a large quantum–classical distance, indicating strong deviation from classical dynamics, can benefit the design of optimized chiral quantum walks~\cite{frigerio2022quantum}.
However, these same features are a critical vulnerability to edge-based QSW decoherence, where they accelerate the decay of quantum coherence. 

Moreover, the value of coherence in a noiseless scenario, particularly in highly connected networks, whose initialization at high-degree nodes tends to promote localization but reduces coherence in the noiseless conditions. This observation highlights a fundamental trade-off between localization and coherence. This trade-off is particularly important for quantum information processing. In applications where coherence is the main resource, such as quantum search algorithms~\cite{shi2017coherence, su2018coherence}, network topologies such as cycle, Erdős–Rényi, and small-world networks are more suitable because they maintain higher coherence. In contrast, for applications that rely on high fidelity and localization, such as robust quantum memories~\cite{chandrashekar2015localized}, star, complete, and scale-free networks are more appropriate.

The present study considers only Markovian single CTQW dynamics on static networks. However, the approach can be extended to include non-Markovian effects~\cite{benedetti2016non}, time-dependent network structures~\cite{PhysRevA.100.012306, PhysRevA.100.062325}, and multiple interacting quantum walks (multi-CTQW)~\cite{doi:10.1126/science.1229957, benedetti2012quantum, zhou2024multi, PhysRevLett.127.100406}. These results can also be tested experimentally, particularly in photonic waveguide arrays and integrated photonic chips~\cite{qu2022experimental, tang2018experimental, zhou2024multi}, where different network structures can be designed and observed.

\section*{Acknowledgments}
ALJ acknowledges the University Grants Commission (UGC), India, for financial support. CM acknowledges support from the Anusandhan National Research Foundation (ANRF) India (Grants Numbers SRG/2023/001846 and EEQ/2023/001080). We are grateful to Dr. Luca Razzoli for helpful correspondence and for pointing us to his work~\cite{razzoli2022universality}, which provided the analytical explanation for the universality we observed in the star initial localization on the center node and complete network dynamics.

\section*{Data availability}
Data sharing is not applicable, as this study did not analyze any raw data.

\appendix
\appendix
\section{Influence of Network Size and Decoherence Rates in Simple Networks}
\label{scaling for simple}
\begin{figure}[htbp]
    \centering
    \includegraphics[width=1\linewidth]{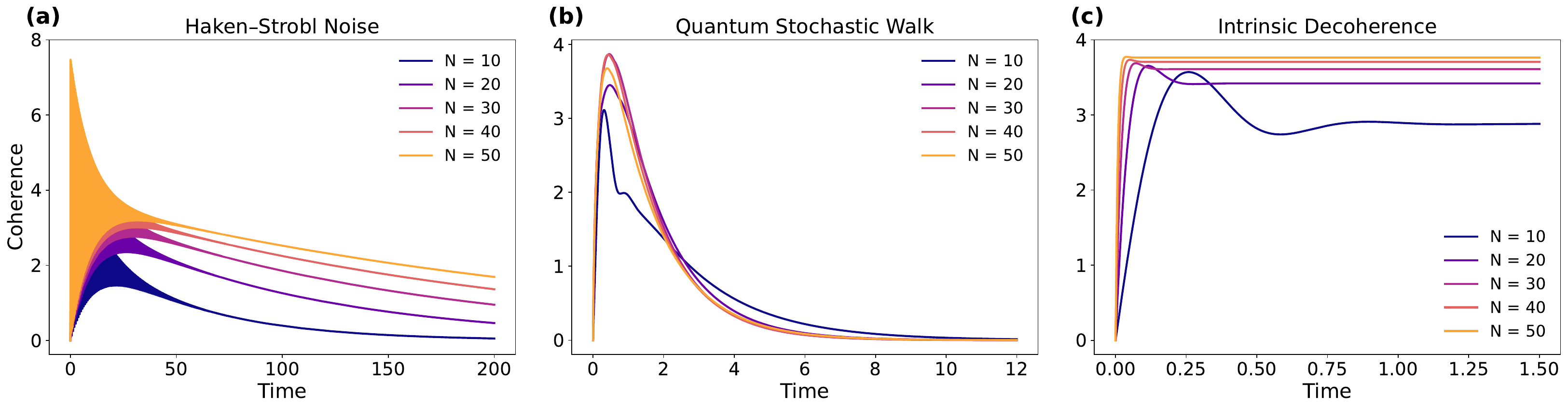}
    \includegraphics[width=1\linewidth]{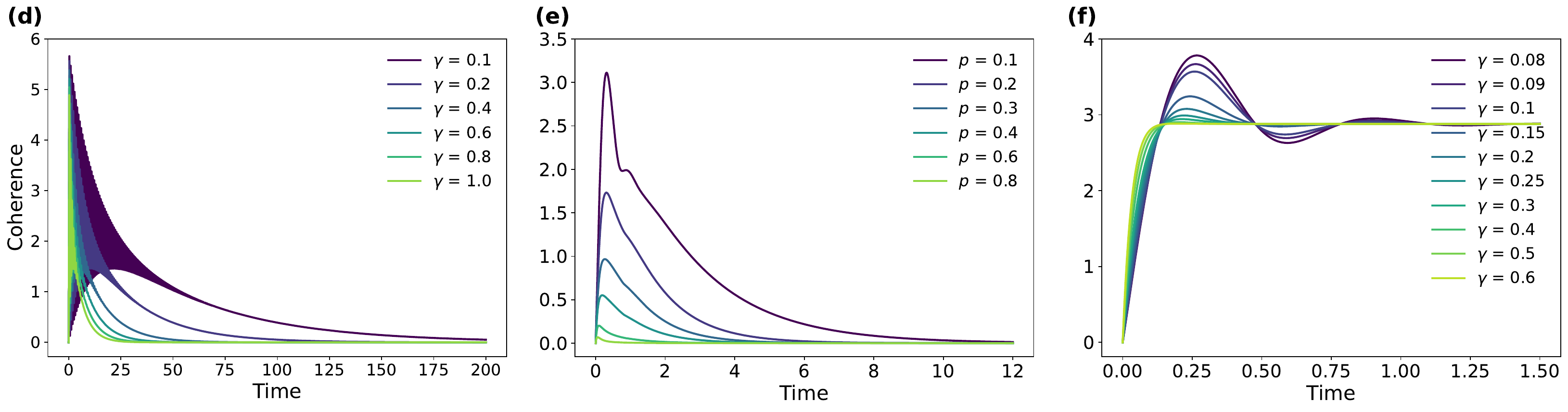}
  \caption{\textbf{Effect of network size and decoherence rate on the $\ell_1$-norm of coherence for a star graph with hub node initialization.} Panels \textcolor{magenta}{(a)–(c)} show the coherence decay for different network sizes $N$ under \textcolor{magenta}{(a)} Haken–Strobl noise, \textcolor{magenta}{(b)} QSW, and \textcolor{magenta}{(c)} intrinsic decoherence, each evaluated at fixed decoherence rates ($\gamma = 0.1$ or $p = 0.1$).
\textcolor{magenta}{(d)–(f)} The bottom row illustrates the coherence decay for various decoherence rates ($\gamma$ or $p$) for the three corresponding decoherence models on a network of fixed size $N = 10$.} 
    \label{fig:coherence_p_gamma}
\end{figure}

This appendix presents a scaling analysis of the star network with central node initialization, examining the dependence of $\ell_1$-norm of coherence on network size $N$ and decoherence rates $\gamma$ and $p$. As shown in Fig.~\ref{fig:coherence_p_gamma}\textcolor{magenta}{(a)}, for Haken-Strobl noise the time taken for coherence decay increases with network size, showing that stability improves as $N$ increases. In contrast, under QSW dynamics, the opposite trend emerges. As observed in Fig.~\ref{fig:coherence_p_gamma}\textcolor{magenta}{(b)}, the coherence decays more rapidly as $N$ increases, indicating that larger networks are slightly less stable under this decoherence mechanism. For intrinsic decoherence, Fig.~\ref{fig:coherence_p_gamma}\textcolor{magenta}{(c)} shows that as $N$ increases, the coherence saturates faster and to a higher steady-state coherence value, suggesting greater stability. Although smaller networks exhibit a slight peak before stabilizing, the reduced overall decay for larger $N$ shows that stability is enhanced. Next, we study the dependence on decoherence rates for a fixed network size of $N=10$. As shown in Fig.~\ref{fig:coherence_p_gamma}\textcolor{magenta}{(d)} for Haken-Strobl noise and Fig.~\ref{fig:coherence_p_gamma}\textcolor{magenta}{(e)} for QSW, stability uniformly decreases as the respective rates $\gamma$ and $p$ increase. For intrinsic decoherence Fig.~\ref{fig:coherence_p_gamma}\textcolor{magenta}{(f)} shows that while larger $\gamma$ value dampen initial oscillations and lead to faster saturation, the final steady-state coherence value is dependent on the network size, not on the decoherence rate.
\section{Proof for Steady state}
\label{proof}
In this appendix we prove that the steady state of both the Haken-Strobl noise and QSW dynamics must relax to the unique maximally mixed state ($\rho_{ss}=\frac{\mathbb{I}}{N}$). This is demonstrated by verifying that their respective Liouvillians are unital and that the systems satisfy the condition for Davies irreducibility~\cite{davies1976quantum}.

\subsection{Steady state for Haken-Strobl Noise}
\label{HSproof}
From Eq.~\ref{eq.HS}, Substituting the Lindblad operators which are projectors $P_{k}= |k\rangle \langle k|$ is given by:
\begin{equation}
        \frac{d\rho(t)}{dt} = \mathcal{L}(\rho)= -i[H, \rho(t)] + \gamma\sum_k \left( |k\rangle \langle k|\rho(t) |k\rangle \langle k| - \frac{1}{2} \{ |k\rangle \langle k |k\rangle \langle k|, \rho(t) \} \right);
        \label{eq.A1}
\end{equation} 
For a steady state, we must have  $\mathcal{L}(\rho^{ss})=0$. We check if the maximally mixed state $\rho^{ss}=c\mathbb{I}$ is a valid steady state:
 \begin{equation}
\begin{split}
\mathcal{L}(c\mathbb{I})
&= -i[H, c\mathbb{I}] 
+ \gamma \sum_k \left( |k\rangle \langle k|\, c\mathbb{I}\, |k\rangle \langle k|
- \frac{1}{2} \left\{ |k\rangle \langle k|k\rangle \langle k|,\, c\mathbb{I} \right\} \right) \\
&= \gamma \sum_k \left( c\,|k\rangle \langle k|
- \frac{1}{2} \left\{ |k\rangle \langle k|,\, c\mathbb{I} \right\} \right)\\&=\sum_{k} c|k\rangle \langle k|-c|k\rangle \langle k|=0.
\end{split}
\label{eq.A2}
\end{equation}Applying the normalization condition $Tr(\rho)=1$. We find that $Nc=1$ or $c=1/N$. Thus, the steady state density matrix is:\begin{equation}
    \rho_{ss}=\frac{\mathbb{I}}{N}
    \label{eq:A4}.
\end{equation}
This shows that ${\frac{\mathbb{I}}{N}}$ is a steady state solution, where N is the number of nodes. 

To guarantee that this steady state is unique, we invoke the concept of Davies irreducibility~\cite{davies1976quantum}. We utilize the criteria proposed by Zhang and Barthel~\cite{zhang2024criteria}, which state that a system is Davies reducible if and only if there exists a non-trivial projection $\mathcal{P}$ where (where ${\mathcal{P}} \neq 0$ and $\mathcal{P} \neq \mathbb{I}$) that satisfies two conditions: (i) $(\mathbb{I} - \mathcal{P}) {P}_k \mathcal{P} = 0$ for all Lindblad operators ${P}_k$.
(ii) $(\mathbb{I} - \mathcal{P})( i{H} + \frac{1}{2}\sum_{k}{P}_{k}^{\dagger}{P}_{k} )\mathcal{P} = 0$. If we can prove that the only projections $\mathcal{P}$ satisfying these conditions are $\mathcal{P} = 0$ and $\mathcal{P} = \mathbb{I}$, the quantum system is irreducible. For the first condition:
\begin{equation}
    (\mathbb{I} - \mathcal{P}) |k\rangle \langle k| \mathcal{P} = 0.
    \label{eq.A5}
\end{equation}
This implies:
\begin{equation}
    |k\rangle \langle k| \mathcal{P}= \mathcal{P}|k\rangle \langle k| \mathcal{P}.
    \label{eq.A6}
\end{equation}Taking hermitian adjoint of this equation Eq.~\eqref{eq.A6} and using the property that projectors $\mathcal{P}$ are hermitian,

\begin{equation}
    \mathcal{P}|k\rangle \langle k| = {P}|k\rangle \langle k|\mathcal{P}.
    \label{eq.A7}
\end{equation}Eq.~\eqref{eq.A6} and Eq.~\eqref{eq.A7} imply that the commutator of $\mathcal{P}$ and $|k\rangle \langle k|$ is zero: $[\mathcal{P},|k\rangle \langle k|]=0$. Evaluating the $ij^{th}$ element of the commutator for $i\neq j$:

\begin{equation}
    \langle i|[\mathcal{P}|k\rangle \langle k|]|j\rangle=\langle i|\mathcal{P}|k\rangle \langle k|j\rangle-\langle i|k\rangle\langle k|\mathcal{P}|j\rangle=\langle i|\mathcal{P}|k\rangle\delta_{kj} -\delta_{ik}\langle k|\mathcal{P}|j\rangle.
    \label{eq.A8}
\end{equation}Let us impose the $\delta_{kj}$ in the first term. Then for $i\neq j$:
\begin{equation}
    \langle i|\mathcal{P}|j\rangle\delta_{jj} -\delta_{ij}\langle j|\mathcal{P}|j\rangle=\langle i|\mathcal{P}|j\rangle-0=\mathcal{P}_{ij}=0.
\end{equation}Since $\langle i|\mathcal{P}|j\rangle =0$ for all $i\neq j$, the projector $\mathcal{P}$ must be diagonal.
For the second condition, substituting $H=L$(the graph Laplacian):
\begin{equation}
    (\mathbb{I} - \mathcal{P})( i{L} + \frac{1}{2}\sum_{k}|k\rangle \langle k|k\rangle \langle k| )\mathcal{P} = (\mathbb{I} - \mathcal{P})( i{L} + \frac{1}{2}\sum_{k}|k\rangle \langle k| )\mathcal{P}=iL\mathcal{P}-iPL\mathcal{P}=0.
    \label{eq.A9}
\end{equation}This implies that $L\mathcal{P}=\mathcal{P}L\mathcal{P}$. Taking adjoint gives $\mathcal{P}L=\mathcal{P}L\mathcal{P}$. Thus, L and $\mathcal{P}$ must commute($[\mathcal{P},L]$=0). Taking $ij^{th}$ element of the commutator,
 \begin{equation}
     [L,\mathcal{P}]_{ij}=\sum_{k} L_{ik}\mathcal{P}_{kj}-\mathcal{P}_{ik}L_{kj}= L_{ij}\mathcal{P}_{jj}-\mathcal{P}_{ii}L_{ij}=L_{ij}(\mathcal{P}_{jj}-\mathcal{P}_{ii}).
      \label{eq.A11}
 \end{equation}Here we used the fact that since $\mathcal{P}$ is diagonal ($\mathcal{P}_{kj}=\mathcal{P}_{kj}\delta_{kj}$). If nodes i and j are connected, $L_{ij}=-1$. Since the commutator is zero, Eq.~\eqref{eq.A11} implies that $P_{jj}= P_{ii}$ for all connected nodes. For a connected graph, this is true for all nodes, meaning all diagonal elements must be equal: $\mathcal{P}=c\mathbb{I}$. 
 
 Since the eigen values of projector can only be 0 or 1, the only possible values are $c= 0$ or $ c=1$. Thus $\mathcal{P}=0$ or $\mathcal{P}=\mathbb{I}$ are the only possible projections. This satisfies the condition for irreducibility. Therefore, the Haken-Strobl Lindblad operators and irreducible and the system has a unique steady state, $\rho_{ss}=\frac{\mathbb{I}}{N}$ as shown in Eq.~\ref{eq:A4}. This analytical result for steady state aligns with our simulations irrespective of network topology as shown in Sec.~\ref{section3a}.
\subsection{Steady state for Quantum Stochastic Walk}
\label{QSWproof}
From Eq.~\ref{eq.QSW}, substituting the Lindblad operators for QSW $P_{kj}= |k\rangle \langle j|$. The master equation is:
\begin{equation}
\begin{split}
\frac{d\rho(t)}{dt} =\mathcal{L}(\rho)= -(1 - p)i[H, \rho(t)] 
 + p \sum_{kj} \left( L_{kj}|k\rangle \langle j| \rho(t) L_{kj}|j\rangle \langle k| 
- \frac{1}{2} \left\{ L_{kj}|j\rangle \langle k| L_{kj}|k\rangle \langle j| \rho(t) \right\} \right);
\end{split}
\label{eq.A12}
\end{equation}

checking whether $c\mathbb{I}$ can be a valid steady state for QSW as done for Haken-Strobl noise~\ref{HSproof}:
\begin{equation}
\begin{split}
\mathcal{L}(c\mathbb{I}) 
&= -(1 - p)i[H, c\mathbb{I}] + p \sum_{kj} \left( L^2_{kj}|k\rangle \langle j|\, c\mathbb{I}\, |j\rangle \langle k|
- \frac{1}{2} \left\{ L^2_{kj}|j\rangle \langle k|k\rangle \langle j| ,\, c\mathbb{I} \right\} \right) \\
&= p \sum_{kj} \left( cL^2_{kj}|k\rangle \langle k|
- \frac{1}{2} \left\{ L^2_{kj}|j\rangle \langle j| ,\, c\mathbb{I} \right\} \right);
\end{split}
\label{eq.A13}
\end{equation}

where, $\frac{1}{2} \left\{ L^2_{kj}|j\rangle \langle j| , c\mathbb{I} \right\} =c L^2_{kj}|j\rangle \langle j| $. Substitute this in Eq.~\ref{eq.A13} gives:
\begin{equation}
    \mathcal{L}(c\mathbb{I})=0.
    \label{eq.A14}
\end{equation}
Since $Tr(\rho^{ss})=1$ gives:$Nc=1$ or $c=1/N$ which means the steady state is:
\begin{equation}
    \rho_{ss}=\frac{\mathbb{I}}{N}.
    \label{eq:A15}
\end{equation}

As we did in ~\ref{HSproof}, we use the Zhang and Barthel criteria to check for irreducibility.
Applying the first condition $(\mathbb{I} - \mathcal{P}) {P}_{kj} \mathcal{P} = 0$:
\begin{equation}
    (\mathbb{I} - \mathcal{P}) L_{kj} |k\rangle \langle j| \mathcal{P} = 0.
    \label{eq:A16}
\end{equation}
Consider the diagonal elements of this equation. $k=j$
\begin{equation}
    (\mathbb{I} - \mathcal{P}) L_{kk} |k\rangle \langle k| \mathcal{P}=0.
    \label{eq:A17}
\end{equation}

This is identical to Eq.~\ref{eq.A5} for the Haken-Strobl case. It immediately implies that for QSW, the projection operator $\mathcal{P}$ must be diagonal.

Now consider the off-diagonal jump operators $k\neq j$. Taking the $il^{th}$ element of this:

\begin{equation}
\begin{split}
\big((\mathbb{I}-\mathcal{P})\,L_{kj}\,|k\rangle\langle j|\,\mathcal{P}\big)_{il}
&= \sum_{mn} L_{kj}\,(\mathbb{I}-\mathcal{P})_{im}\,(|k\rangle\langle j|)_{mn}\,\mathcal{P}_{nl} \\
&= \sum_{mn} L_{kj}\,(\mathbb{I}-\mathcal{P})_{im}\,\delta_{m k}\,\delta_{n j}\,\mathcal{P}_{nl} \\
&= L_{kj}\,(\mathbb{I}-\mathcal{P})_{i k}\,\mathcal{P}_{j l} \\
&= L_{kj}\,\big(1 - \mathcal{P}_{kk}\big)\,\mathcal{P}_{j l}\,.
\end{split}
\label{eq.A18}
\end{equation}
Here we used the result that $\mathcal{P}$ should be  diagonal from Eq.~\ref{eq:A17}.
Also by Eq.~\eqref{eq:A16},
$L_{kj}(1-\mathcal{P}_{kk})\mathcal{P}_{jj}=0$. If node i and j are connected, the Laplacian elements $L_{kj}=-1$ else $L_{kj}=0$. Therefore, for connected nodes k and j: (i) If $\mathcal{P}_{jj}=1$ (node j is in the subspace), then it forces $(1-\mathcal{P}_{kk})=0$, so $\mathcal{P}_{kk}=1$ (node k is also in the same subspace). (ii) If $\mathcal{P}_{kk}=0$ (node k is out of the subspace), then $\mathcal{P}_{jj}=0$.

Since the graph is connected, either all diagonal elements are 1 ($\mathcal{P}=\mathbb{I}$) or all are 0 ($\mathcal{P}=0$). This proves that the QSW dynamics is Davies irreducible and has a unique steady state as in Eq.~\eqref{eq:A15}, $\rho^{ss}=\frac{\mathbb{I}}{N}$. This analytical result perfectly aligns with our numerical simulations in Sec.~\ref{section3a} and ~\ref{section3b}, where we consistently observed convergence to this uniform distribution of probabilities with coherence decays to zero regardless of the network topology.

\vspace{5pt}

\section{Linking Initialized Node's Network Properties with Relaxation Time}\label{sec.network_properies_and_tau}

\begin{figure}[ht]
    \centering
    \includegraphics[width=1\linewidth]{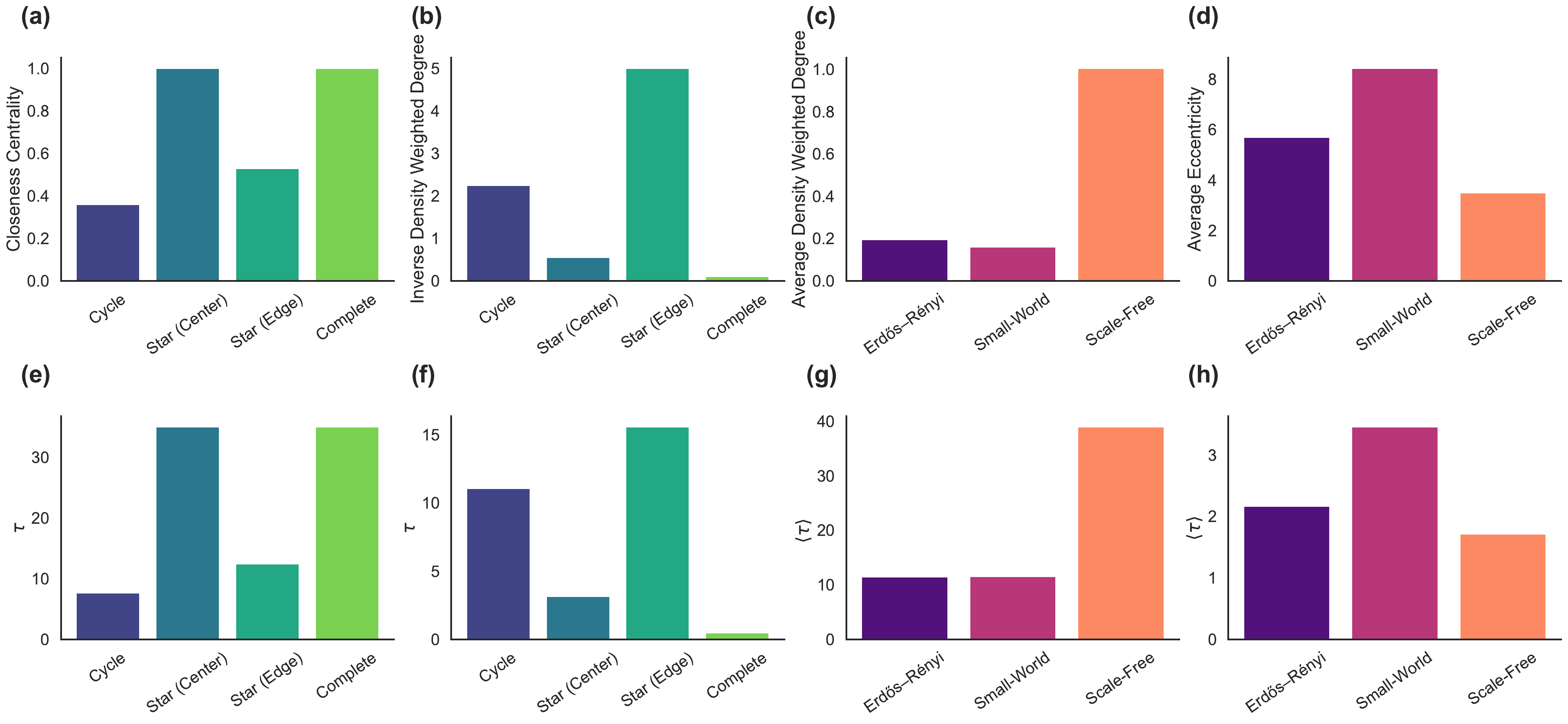}
      \caption{\textbf{Comparison of CTQW initialized node's properties and relaxation time hierarchies.} Panels \textcolor{magenta}{(a)–(d)} display specific properties of initialized node, which are compared against the characteristic relaxation times $\tau$ shown in panels \textcolor{magenta}{(e)-(h)}. Specifically, panels \textcolor{magenta}{(e)} and \textcolor{magenta}{(f)} show the relaxation times for simple topologies (Table~\ref{tab:coherence_decay}), while panels \textcolor{magenta}{(g)} and \textcolor{magenta}{(h)} correspond to the complex network ensembles (Table~\ref{tab:network_decay}) under Haken-Strobl and QSW noise, respectively.}
    \label{fig:network_properties}
\end{figure}

In this Appendix, we illustrate the relationship between properties of initialized node and the characteristic relaxation time $\tau$ (Eq.~\eqref{eq:tau}), which is derived from the Kohlrausch parameters $\beta$ and $\lambda$ in Eq.~\eqref{decay_fit}.  

For simple network topologies (cycle, star, and complete), the closeness centrality of the initialization node (Fig.~\ref{fig:network_properties}\textcolor{magenta}{(a)}) follows the same hierarchy as the $\tau$ values observed under Haken-Strobl noise Fig.~\ref{fig:network_properties}\textcolor{magenta}{(e)} and Table~\ref{tab:coherence_decay}. In the QSW regime, however, the relaxation times in Fig.~\ref{fig:network_properties}\textcolor{magenta}{(f)} align with the inverse density-weighted degree of the initial node $\frac{1}{k_i\rho}$ , as shown in Fig.~\ref{fig:network_properties}\textcolor{magenta}{(b)}. Here, $k_i$ represents the degree of the initial node and $\rho=\frac{2E}{N(N-1)}$ denotes the network density~\cite{BEDRU2020100247}, where E and N are the total number of edges and nodes, respectively.

For complex network ensembles, these node properties are ensemble-averaged over the same realizations used to compute the relaxation times in Table.~\ref{tab:network_decay}.
We observe that the ensemble average of density-weighted degree of the initial node ($\langle k_i\rho\rangle$)  as shown in Fig.~\ref{fig:network_properties}\textcolor{magenta}{(c)} mirrors the hierarchy of $\langle \tau \rangle$ under Haken-Strobl noise (Fig.~\ref{fig:network_properties}\textcolor{magenta}{(g)}). Conversely, for QSW decoherence, the average eccentricity of the initial node Fig.~\ref{fig:network_properties}\textcolor{magenta}{(d)} aligns well with the behavior of $\langle \tau \rangle$ (Fig.~\ref{fig:network_properties}\textcolor{magenta}{(h)}), further highlighting how specific initialized node properties capture the stability of the system under different decoherence mechanisms.

This hierarchy dictates how the properties of initially localized node affects the retention of coherence. Under Haken-Strobl noise, a node based decoherence mechanism, high closeness centrality and a high average density weighted degree shows high relaxation times, indicating that well-connected starting node can sustain coherence for longer. In contrast under QSW decoherence, which depends on both nodes and edges, the highest relaxation correlates with inverse of density weighted degree and eccentricity of the initial node. This suggests that nodes which posses fewer connectivity or topologically remote retain coherence for longer durations.

\section{Extended Stability Metrics for Simple Topologies}\label{sec.complete_cycle}

\begin{figure}[ht]
    \centering

        \centering
        \includegraphics[width=\textwidth]{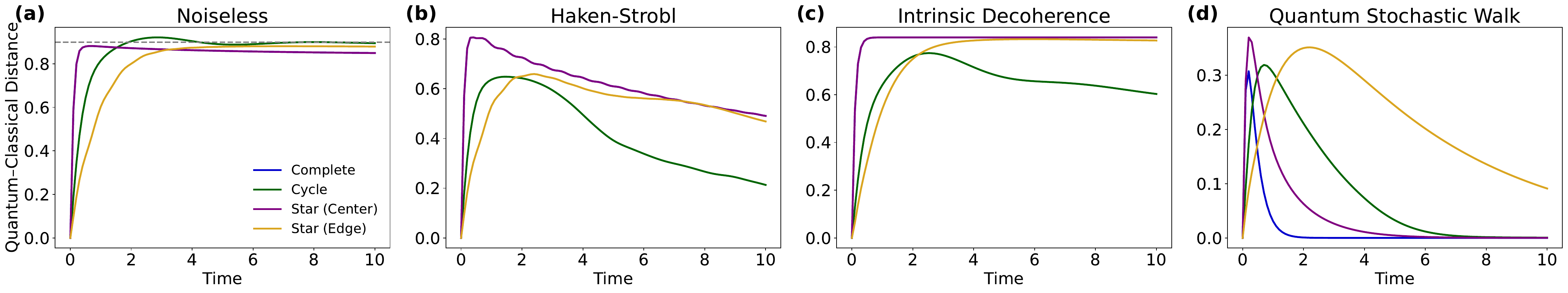}

        \centering
        \includegraphics[width=\textwidth]{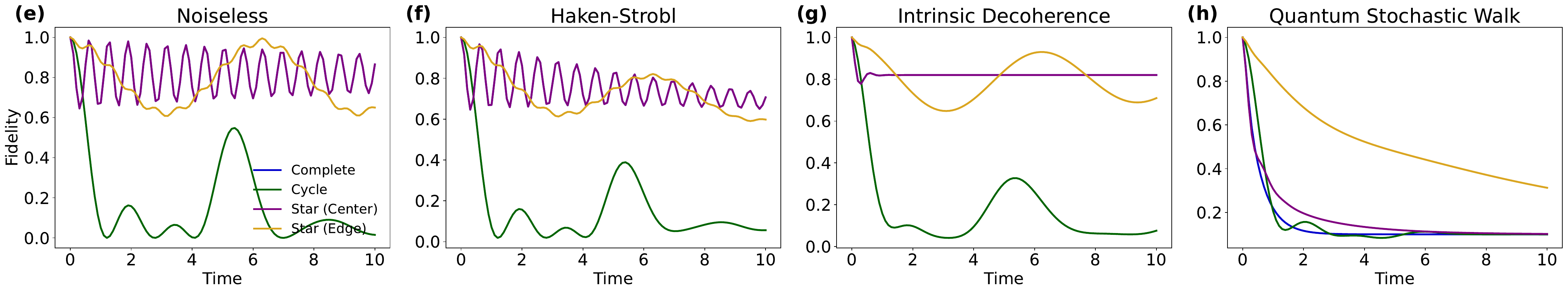}
    
    \caption{\textbf{Quantum-classical distance and fidelity with the initial state for complete, cycle and star networks with size N = 10}. Quantum-classical distance and fidelity with the initial state for noisless \textcolor{magenta}{(a),(e)} and three decoherence models: Haken-Strobl \textcolor{magenta}{(b),(f)}, intrinsic Decoherence \textcolor{magenta}{(c),(g)} , and QSW \textcolor{magenta}{(d),(h)}. The grey dashed line in \textcolor{magenta}{(a)} is theoretical value of quantum classical distance under no noise.}
    \label{fig:qc_fid_combined_2}
\end{figure}

\begin{figure}[ht]
    \centering
    \includegraphics[width=0.75\textwidth]{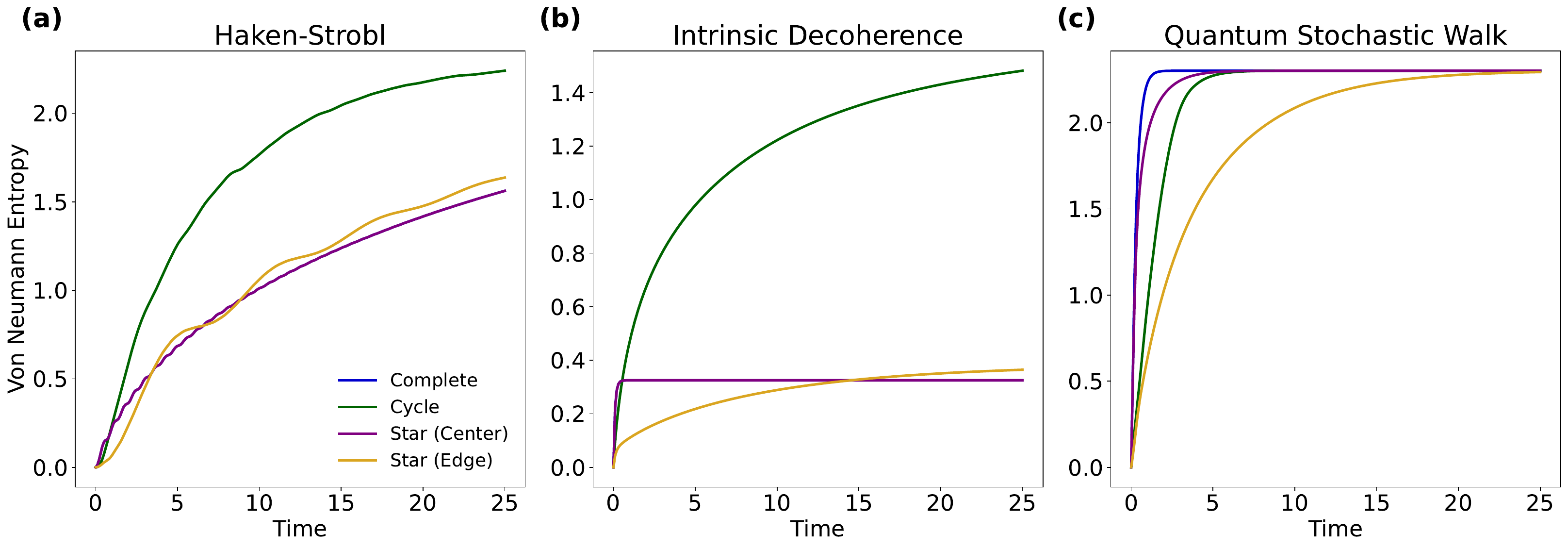}
    \caption{\textbf{Time evolution of von Neumann entropy for Complete, Cycle and Star networks of size N = 10}. Von Neumann entropy under Haken-Strobl \textcolor{magenta}{(a)}, Intrinsic Decoherence \textcolor{magenta}{(b)}, and QSW dechorence \textcolor{magenta}{(c)}.}
    \label{fig:14_2}
\end{figure}

To obtain a more comprehensive picture of stability beyond the $\ell_1$-norm of coherence in Sec.~\ref{section3a}, we also analyze the quantum-classical distance, fidelity with the initial state, and von Neumann entropy. Indeed, these additional metrics confirm that the complete graph and the star network (hub-initialized) consistently demonstrate the highest stability, a finding that holds for both Haken-Strobl and intrinsic decoherence. This reinforces our earlier conclusion from the $\ell_1$-norm of coherence analysis, which had already identified these  two same networks with high degree nodes as the most robust under the Haken-Strobl model. This is evidenced by their high quantum-classical distance and fidelity, along with the lowest von Neumann entropy, as shown in Fig.~\ref{fig:qc_fid_combined_2}\textcolor{magenta}{(a),(b),(c),(e),(f),(g)} and Fig.~\ref{fig:14_2}\textcolor{magenta}{(a),(b)}. In contrast, the cycle network tends to classicalize more rapidly, indicated by its low quantum-classical distance (Fig.~\ref{fig:qc_fid_combined_2}\textcolor{magenta}{(b),(c)}). 
However, these metrics also reveal that for the QSW model, the stability rankings invert. The cycle network, previously the least stable, now displays a slower classicalization (a higher quantum calssical distance) than the highly-connected complete and star network with center node initialization (Fig.~\ref{fig:qc_fid_combined_2}\textcolor{magenta}{(d)}). Under QSW, the star network with peripheral node initialization emerges as the most stable configuration, showing the slowest decay of quantum-classical distance, high fidelity, and low entropy (Fig.~\ref{fig:qc_fid_combined_2}\textcolor{magenta}{(d),(h)} and Fig.~\ref{fig:14_2}\textcolor{magenta}{(c)}). This is consistent with our coherence analysis, which also identified this configuration as the most robust under QSW. This emphasizes the intricate interplay between network topology, the walker's initial condition, and the specific mechanism of decoherence, as elaborated on in the main conclusion (Sec.~\ref{section5}).

\section{Influence of Average Degree and Network Size in Complex Networks}
\label{averagre}

In this appendix, we investigate the impact of the average degree $\langle k \rangle$ and network size $N$ on the temporal evolution of coherence across different network topologies. 

\begin{figure}[htbp]
\includegraphics[width=\textwidth]{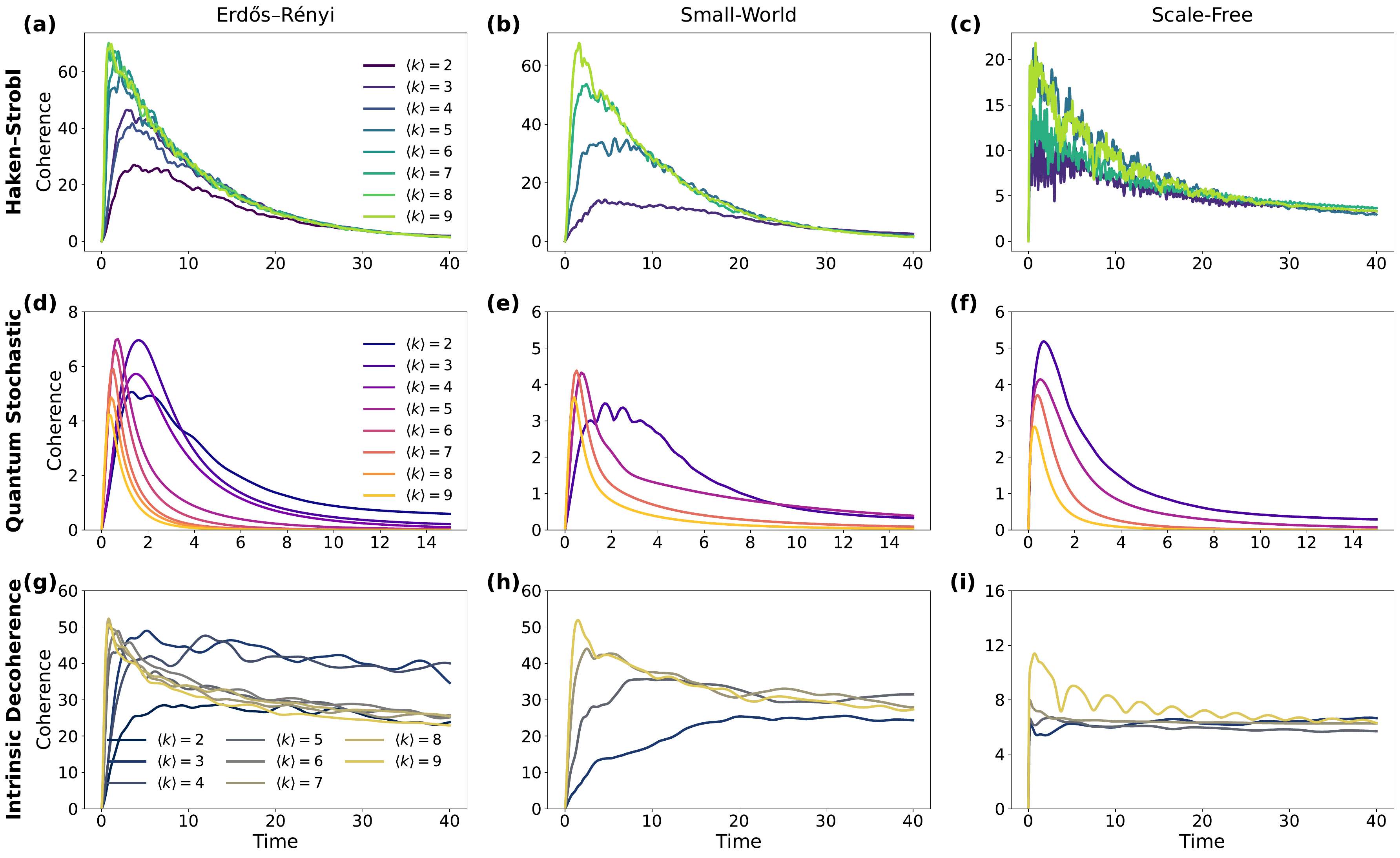}
    \caption{\textbf{Influence of average degree $\langle k \rangle$ on $\ell_1$-norm coherence in complex networks.}Temporal evolution of coherence for Haken-Strobl noise in \textcolor{magenta}{(a)-(c)}, QSW decoherence in \textcolor{magenta}{(d)-(f)}, and intrinsic decoherence in \textcolor{magenta}{(g)-(i)} across Erdős--Rényi, small-world, and scale-free topologies ($N=100$). For the small-world (Watts–Strogatz) and scale-free (Barabási–Albert) models, only even values of $\langle k \rangle$ are considered to satisfy the generation constraints of these models.}
    \label{Fig:average_degree}
\end{figure}

To evaluate the effect of average degree, we examine the $\ell_1$-norm of coherence across a range of its values. As shown in Fig.~\ref{Fig:average_degree}\textcolor{magenta}{(a)-(c)}, the peak magnitude of the $\ell_1$-norm of coherence under Haken-Strobl noise increases with $\langle k \rangle$ across Erdős–Rényi and small-world topologies, a trend that holds for most average degree values in scale-free networks as well. Notably, as  $\langle k \rangle$ increases  decay of coherence becomes more rapid for QSW decoherence for all network topologies as shown in Fig.~\ref{Fig:average_degree}\textcolor{magenta}{(d)-(f)}. Furthermore, the intrinsic decoherence model depicted in Fig.~\ref{Fig:average_degree}\textcolor{magenta}{(g)-(i)} reveals a distinct qualitative shift in dynamics; while coherence saturates to a stable value for low connectivity ($\langle k \rangle \lessapprox 4$), systems with higher connectivity ($\langle k \rangle \gtrapprox 5$) exhibit a transient maximum followed by a decay toward a non-trivial constant steady state $\ell_1$-norm of coherence.

\begin{figure}[htbp]
\includegraphics[width=\textwidth]{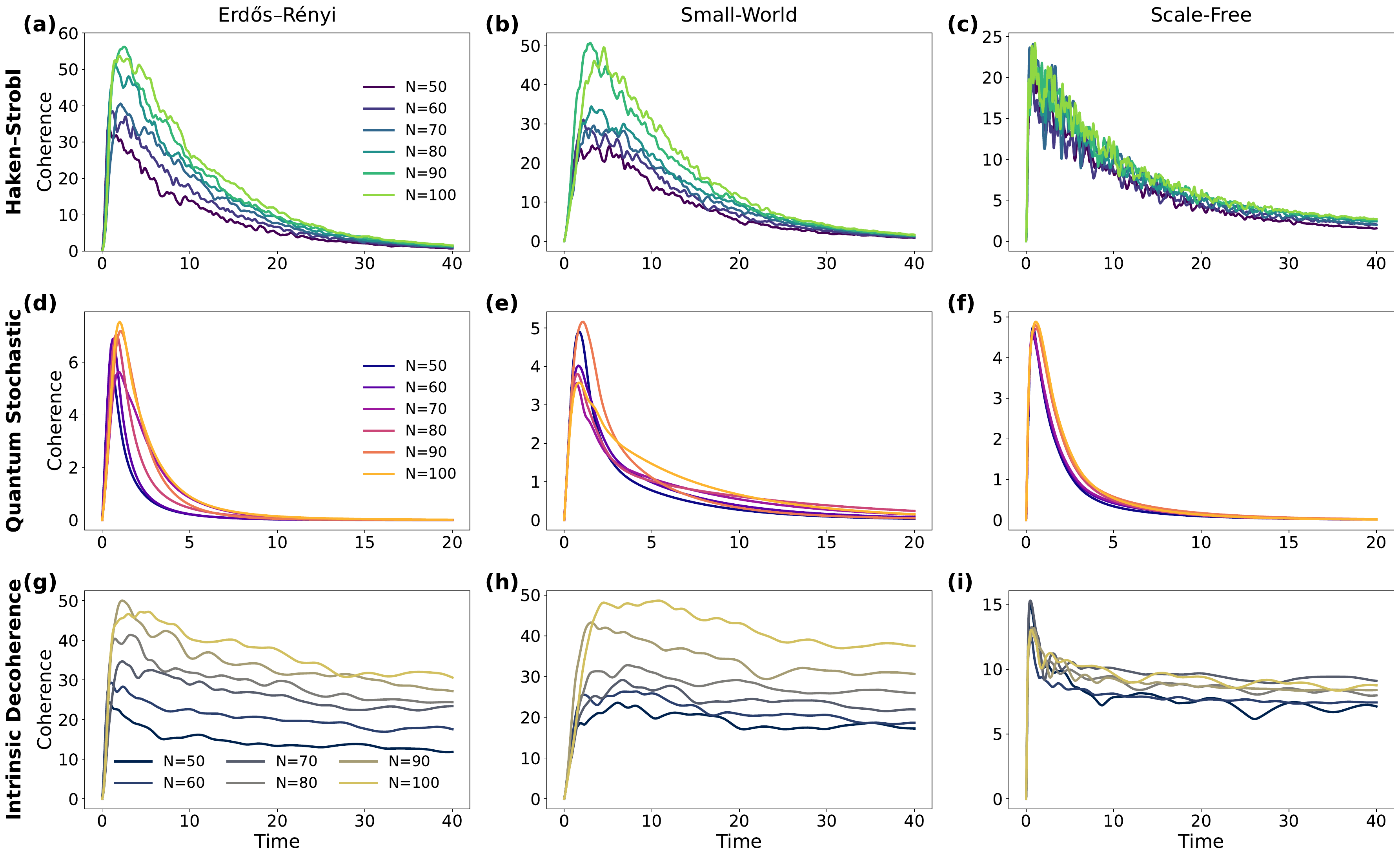}
    \caption{\textbf{Influence of network size $N$ on $\ell_1$-norm coherence in complex networks.} The temporal evolution of coherence is presented for Haken-Strobl noise in \textcolor{magenta}{(a)-(c)}, QSW decoherence in \textcolor{magenta}{(d)-(f)}, and intrinsic decoherence in \textcolor{magenta}{(g)-(i)} across Erdős--Rényi, small-world, and scale-free topologies with average degree $\langle k \rangle =4$.}
    \label{Fig:complex_size}
\end{figure}

The influence of network size $N$ on complex topologies is presented in Fig.~\ref{Fig:complex_size}. Under Haken-Strobl noise (Fig.~\ref{Fig:complex_size}\textcolor{magenta}{(a)-(c)}), ncreasing $N$ leads to higher peak coherence and more rapid decay rates in Erdős–Rényi and small-world networks; however, scale-free networks exhibit relative invariance to system size, likely due to the dominant role of the central hub. For QSW decoherence (Fig.~\ref{Fig:complex_size}\textcolor{magenta}{(d)-(f)}), the coherence trajectories display a less discernible dependence on $N$, suggesting that coherence decay is primarily governed by local connectivity rather than global system size. In contrast, under intrinsic decoherence (Fig.~\ref{Fig:complex_size}\textcolor{magenta}{(g)-(i)}), a clear scaling trend emerges: the  value at which coherence stabilizes, whether attained via direct saturation or following a transient peak, scales positively with $N$ for Erdős–Rényi and small-world topologies. An analogous trend observed in scale-free networks across some of the investigated system sizes ($N=50,60,70$).

\section{Node Centrality Effects in Scale-Free Networks}
\label{initial_scalefree}
\begin{figure}[ht]
\includegraphics[width=1\textwidth]{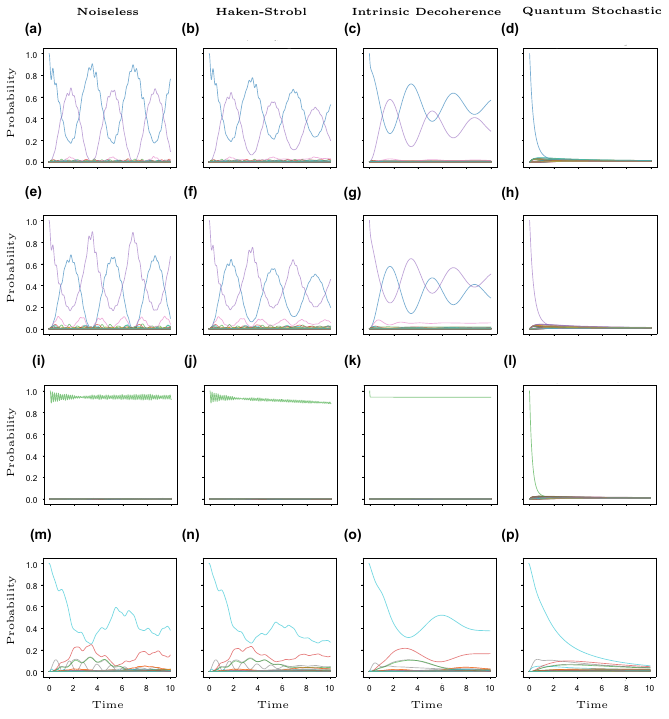}
    \caption{\textbf{Node occupation probability versus time under different initial conditions of scale-free networks with size $N = 100$}
\textcolor{magenta}{(a)–(d)}: The network structure has distinct nodes for the highest degree and highest closeness centrality. The walker is initially localized on the highest-degree node, which exhibits anti-synchronization with highest-closeness node.
\textcolor{magenta}{(e)–(h)}: Similar network configuration as above, but the walker is initially localized on the highest-closeness node. anti-synchronization is observed with the highest-degree node in this case.
\textcolor{magenta}{(i)–(l)}: The walker starts at a node where both the highest degree and highest closeness centralities coincide.
\textcolor{magenta}{(m)–(p)}: The walker is initialized on the lowest-degree node. 
    }
\label{fig:10}
\end{figure}
In scale-free networks, nodes often exhibit significant variability in centrality measures such as degree and closeness. We find that when the nodes with the highest degree and highest closeness centrality are distinct, the dynamics of the CTQW exhibit a striking anti-synchronization pattern. Specifically, when the walker is initialized on either the highest-degree node (Fig.\ref{fig:10}\textcolor{magenta}{(a)}) or the highest-closeness node (Fig.\ref{fig:10}\textcolor{magenta}{(b)}), the probability distributions over time show out-of-phase oscillations between these two nodes, suggesting a dynamical competition in occupation probabilities.

In contrast, when the highest-degree and highest-closeness centrality coincide at a single node, the walker remains strongly localized on the initial node throughout the evolution, as shown in Fig.~\ref{fig:10}\textcolor{magenta}{(c)}. Even in the presence of Haken–Strobl decoherence, the decay of the occupation probability is significantly slower, indicating a pronounced robustness of localization that resists noise-induced spreading.

On the other hand, when the walker is initialized on the lowest-degree node (Fig.~\ref{fig:10}\textcolor{magenta}{(d)}), the CTQW exhibits no such localization behavior. The walker spreads more readily across the network, and the occupation probability of the initial node diminishes rapidly. Under quantum stochastic walk dynamics, the probability distribution in this case reaches equilibration more slowly compared to scenarios involving high-degree or high-closeness nodes. These findings underscore the critical role of node centrality in shaping both coherence preservation and transport behavior in heterogeneous quantum networks.

\clearpage

\bibliographystyle{unsrt}
\bibliography{reference}
\end{document}